\documentclass{elsart}
\usepackage[dvips]{graphicx}
\usepackage{amsmath}

\begin{document}

\bibliographystyle{plain}

\begin{frontmatter}

\title{
Electromagnetic Coulomb Gas
 with Vector Charges and ``Elastic'' Potentials  : 
Renormalization Group Equations
}

\author{David Carpentier}
\address{ Laboratoire de Physique
de l'Ecole Normale Sup{\'e}rieure de Lyon - CNRS,\\
          46, All{\'e}e d'Italie, 69364 Lyon Cedex 07, France}
\author{Pierre Le Doussal}
\address{ Laboratoire de
Physique Th{\'e}orique de l'Ecole Normale Sup{\'e}rieure - CNRS,\\
          24, Rue Lhomond 75005 Paris, France}

\begin{abstract}
 We present a detailed derivation of the renormalization group equations for two dimensional electromagnetic
Coulomb gases whose charges lie on a triangular lattice (magnetic charges) and its dual (electric charges). The interactions between the charges involve both angular couplings and a new electromagnetic
potential. This motivates
the denomination of ``elastic'' Coulomb gas. 
 Such elastic Coulomb gases arise naturally in the study of the continuous melting transition of two dimensional solids coupled to a substrate, either commensurate or with quenched disorder. 
\end{abstract}

\end{frontmatter}
\section{Introduction}

 The understanding of defect-mediated phase transitions in two dimensions relies on the renormalization group 
 study of Coulomb gases (CG). In the simplest examples of the $O(2)$ or XY model, the criticality of the Kosterlitz-Thouless phase transition is described using the scalar Coulomb Gas\cite{kosterlitz73}. In this case, the charges correspond to the integer topological charges of the XY vortices, which interact {\it via} the 2D Coulomb ($\ln$) potential.  
  If we perturb the  XY model by a $p$-fold  symmetry breaking potential (the so called clock model), the previous scalar 
  CG has to be extended : the clock potential translates into magnetic scalar charges\cite{jose77}. These magnetic 
  charges mutually interact {\it via} the same Coulomb potential, and  their coupling with electric charges  
is a Aharonov-Bohm potential \cite{kadanoff78}. This scalar electromagnetic CG has been studied using the real-space 
renormalization techniques \cite{jose77,nienhuis87}, which provide the critical properties of the initial clock model. 
    Moreover, the phase transitions of various two dimensional models, such as the Ashkin-Teller model, the q-state 
Potts model, and the O(n) model can be studied using these scalar CG techniques\cite{kadanoff78,nienhuis87}. 
  
   An extension of the scalar (electric) Coulomb gas is required in the study of the continuous melting transition of 
a two dimensional solid\cite{nelson83b}. This extension is twofold : (i) the topological charges of two dimensional dislocations are Burgers 
vectors instead of integers and 
(ii) the interaction between these vector charges consist of the usual 2D Coulomb potential ({\it i.e} $\ln$ interaction), 
 and an {\it angular interaction} which couples the charges to the vector ${\bf r}_{12}$ joining the two defects\footnote{ 
 This angular interaction is a manifestation of the microscopic nature of the dislocations, which can be viewed as 
 additional half-line of atoms inserted in the lattice\cite{landau}. A pair of dislocations of opposite Burgers vectors, which 
 is an extra segment of atoms, has obviously some preferred orientation with respect to the initial regular lattice. }.
This angular interaction spoils the conformal invariance of the $\ln$ CG. The renormalization 
group study of the conformally invariant case was achieved in ref.~\cite{nelson78}. 
 Studying the melting transition in the general case amounts to consider the perturbation by marginal conformal (rotation) symmetry-breaking operators of the previous conformal fixed point. The study of the corresponding vector CG was performed in 
 \cite{nelson79,young79}. The natural extension of this vector CG to the electromagnetic case arises in the study of two 
 dimensional melting in the presence of a translation symmetry breaking potential, {\it e.g} a coupling to a substrate via a periodic modulation of the density, as in Ref. \cite{nelson79}. 
 Such a general vector electromagnetic CG has never been studied to our knowledge, and it is the purpose of the present paper to derive the 
 RG equations describing its scaling behavior to lowest order. A preliminary study, motivated by
the problem of a substrate with quenched disorder \cite{carpentier97} was published some time ago, and
involved a replicated VECG \cite{carpentier98a}. The present
study provides a complete and general derivation of the RG equations valid for any
type of substrate (periodic and/or disordered). 
The VECG studied here can be viewed as an extension 
 to the vector/elastic case of the scalar electromagnetic CG \cite{nienhuis87}, and an extension to the electromagnetic case 
 of the vector CG of \cite{nelson79,young79}. As we will see, the elasticity manifests itself not only in the angular interactions 
 of the electric/electric and magnetic/magnetic potentials, but also into the electric/magnetic interaction which is no longer a 
 simpler Aharonov-Bohm potential. 
 
 Before turning to a more precise definition of our model, let us mention  the field theoretical approach to the CG problem. The scalar electromagnetic CG admits an 
equivalent Sine-Gordon field theoretical formulation \cite{wiegmann78}.
Its scaling behavior in the electric case was derived in Ref. \cite{amit80}.  Extension to the electromagnetic CG case were considered 
 in  \cite{boyanovsky89,boyanovsky90} (see also \cite{bulgadaev81}), 
 which included in particular parafermionic operators\cite{fradkin80}. This electromagnetic 
 $\ln-$CG was extended to consider charges in higher groups\cite{boyanovsky91}, as well as relations with string theory models. 
 In these generalized Toda field theories, 
 the CG charges appear as root vectors of Lie algebra, and the charges of the $SU(3)$ Toda field theory can be identified with 
 Burgers vectors of a triangular lattice.  In this perspective, our present study corresponds to an extension to the non-conformal case where angular interactions are included
 of the $SU(3)$ study of Boyanovsky and Holman . 

 %
%
%

  The paper is organized as follows : in section \ref{sec:models}, we derive the CG formulation of an elastic solid coupled to a substrate. 
  We consider explicitly two important cases : the case of a periodic commensurate substrate, and the case of a random pinning 
  substrate. This allows to define the general vector ``elastic'' CG which is the subject of this paper. In section \ref{sec:renormalisation}, 
  the renormalization group equations for this general CG are derived to order one loop, using a real space procedure similar 
  in spirit to the method described in \cite{nienhuis87}. The results are summarized in
Section \ref{sec:RG-general}. Finally, in section \ref{sec:RG-elastic} these equations are restricted to 
the original elastic models. Due to the complexity of the present derivation we have 
deferred to a separate publication the study of these RG equations for the various models. 

\subsection*{Notations}

Throughout this paper, we  use the notations 
$\int_{\vec{r}}=\int d^{2}\vec{r}=\int_{-\infty}^{+\infty}dx dy $ and  
$\int_{\vec{q}}=\int d^{2}\vec{q} / (2\pi)^{2}$. The notation
$\vec{r}$ corresponds to vectors in the two dimensional plane,
originating from either the direct or dual lattice, while boldfaces
$\mathbf{A}$ denote vectors in the replica space. Vectors both
in replica and two dimensional plane $A_{i,a}$ are denoted 
$\vec {\bf A}$. The sum over repeated (real space or replica) 
indices will be assumed :   
\begin{equation}
 A_{i,a} B_{i,a} = \sum_{i=1,2}\sum_{a=1}^{n} A_{i,a} B_{i,a}
\end{equation}
and we use the convolution notation 
\begin{equation}
[A*B](\vec{r}) = \int_{\vec{r}'} A(\vec{r}') B(\vec{r}-\vec{r}')    
\end{equation}
 which for a density of charges
$\vec{b}(\vec{r})=
\sum_{\alpha}\vec{b}_{\alpha} \delta(\vec{r}-\vec{r}_{\alpha})$ reduces to 
\begin{equation}
 b_{i}*V_{ij}*b_{j} = \sum_{i,j=1,2} \sum_{\alpha \beta} 
b_{\alpha,i}
V_{ij}(\vec{r}_{\alpha}-\vec{r}_{\beta}) 
b_{\beta,j}
\end{equation}
Unless otherwise stated, the indices $i,j,k,l$ will correspond to
real space indices $i=1,2$; $a,b,c,d$ to replica indices between $1$
and $n$; and greek indices $\alpha,\beta$ label the charges in a collection
of charges. The notation $\hat{e}$ corresponds to the unit vector
$\vec{e}/|\vec{e}|$.

\section{The model}\label{sec:models}

\subsection{Elastic description of a pinned two dimensional
crystal}\label{sec:elasticity}

\subsubsection{Two dimensional elastic energy}\label{sec:Helastic}

\begin{figure}
\centerline{\includegraphics[width=7cm]{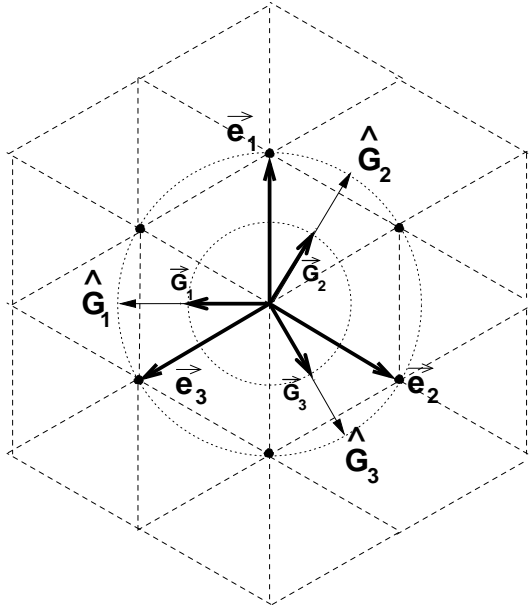}}
\caption{Representation of a hexagonal lattice we will consider in
this paper. The 6 vectors $\pm \vec{e}_{i=1,2,3}$ are the unit vectors
of the original lattices (here the lattice spacing has been set to
$a_{0}=1$), and the 6 unit vectors $\pm \hat{G}_{i=1,2,3}$ lies on the
dual lattice.}
\label{fig:hexagon}
\end{figure}

 In this paper, we will consider a crystal with hexagonal symetry
(see Fig.\ref{fig:hexagon}). For such a lattice, the elasticity is
isotropic, 
and the elastic energy  is given 
by the harmonic hamiltonian\cite{landau}

\begin{align}
\label{eq:H-elastic}
H_0[\vec u] &= 
\frac{1}{2} \int d^2\vec{r}~ u_{ij}(\vec{r}) C_{ijkl} u_{kl}(\vec{r}) = 
\frac{1}{2} \int d^2\vec{r} \left( 2 \mu u_{ij}^2 + 
 \lambda u_{kk}^2 \right) 
\\
& = \frac{1}{2} \int \frac{d^2\vec{q}} 
 {(2\pi)^{2}} ~u_i(\vec{q}) \Phi_{ij}(\vec{q}) u_j(-\vec{q})
\end{align}
 with $C_{ijkl}=\mu (\delta_{ik}\delta_{jl}+\delta_{il}\delta_{jk})
+\lambda \delta_{ij}\delta_{kl}$ 
where $\lambda,\mu$ are Lam{\'e} coefficients, and the tensor 
$u_{ij}$ is defined by\footnote{Note that we have neglected the nonlinear
component of $u_{ij}$.}
$u_{ij}=\frac{1}{2} (\partial_i u_j  +\partial_j u_i )$.  
  For later convenience, it is useful to define the local stress
tensor $\sigma_{ij}=C_{ijkl}u_{kl}=2\mu
u_{ij}+\lambda\delta_{ij}u_{kk}$. 
The elastic matrix $\Phi_{ij}(\vec{q})$ is given by 
\begin{equation}\label{eq:Phi}
\Phi_{ij} (\vec{q} )=
c_{11} q^{2} P_{ij}^{L}(\vec{q}) 
+c_{66}q^{2}P_{ij}^{T}(\vec{q})
\quad ;\quad 
c_{11}=2\mu+\lambda \ ; c_{66}=\mu
\end{equation}
where $c_{11},c_{66}$ are respectively the compression and shear
modulii, and $\lambda,\mu$ the Lam{\'e} coefficients of the crystal. 
We have used the projectors
$P_{ij}^{L}(\vec{q})=\hat{q}_{i}\hat{q}_{j}$,  
$P_{ij}^{T}(\vec{q})=\hat{q}_{i}^{\perp}\hat{q}_{j}^{\perp}
=\delta_{ij}-\hat{q}_{i}\hat{q}_{j}$. 
In this expression, $\vec{u} $ is a smooth displacement field, which
corresponds to the long wavelength distortions of the original
lattice.  Within the context of elasticity, it must satisfy the
condition $|\vec{u}(\vec{r} )-\vec{u} (\vec{r} +\vec{e}_{i} )|\ll a_{0} $,
where $\vec{e}_{i} $ is one of the unit vectors of the original
lattice, and $a_{0}$ the lattice spacing.
  To go beyond this elastic description of the lattice distortions, 
one must allow for dislocations, which are the
topological excitations of this elastic model.

\subsubsection{Two dimensional dislocations}\label{sec:dislocations}

 A two dimensional (edge) dislocation located in $\vec{r} _{\alpha}$ is 
characterised by its topological charge called the Burgers vector 
$\vec{b}_{\alpha}$. This Burgers vector lies on the original lattice,
and for most of our purpose, we will restrict ourselves to unit Burgers
vectors corresponding to one of the six 
$\vec{e}_{i}, i=1,\dots 6$. By definition,
this Burgers vector corresponds to the increment of the displacement
field when surrounding the dislocation : 
\begin{equation}
\oint \vec{u}(\vec{r})~ dl = a_{0}\vec{b}   
\end{equation}
 where the contour integral circles around $\vec{r} _{\alpha}$, and we
choose to consider dimensionless Burgers vectors $\vec{b}$. 
A collection of dislocations can be described by the Burgers vector
density 
\begin{equation}
\vec{b} (\vec{r} )=
\sum_{\alpha}\vec{b} _{\alpha} \delta (\vec{r}-\vec{r}_{\alpha})
\end{equation}
 This density of dislocations 
induces a density of strain relaxed by a displacement field
 $\vec{u}_{d} (\vec{r} )$, derived in appendix \ref{app:dislocations},
and given by\cite{shi91}
\begin{align}\label{eq:u_d}
 u_{d,i}(\vec{r}) & =  \frac{a_{0}}{2\pi}~ 
\left[ \mathcal{G}_{ij}* b_j \right](\vec{r})
= \frac{a_{0}}{2\pi}~
 \sum_{\alpha}  \tilde{\mathcal{G}}_{ij} (\vec{r}-\vec{r}_{\alpha}) ~b_{\alpha ,j}\\
&\textrm{ with }
\tilde{\mathcal{G}}_{ij}(\vec{r})  = 
\delta_{ij}\Phi (\vec{r})
+\frac{c_{66}}{ c_{11}} \epsilon_{ij} \tilde{G}(r) 
+\frac{c_{11} - c_{66}}{c_{11}} \epsilon_{jk} 
 H_{ik}(\vec{r}) 
\end{align}
 The potential $\Phi(\vec{r})$ gives the angle between the vector 
$\vec{r}$ and {\it e.g} the $\vec{e}_{1}$ vector, $\tilde{G}(\vec{r})$
corresponds to the usual ({\it e.g} lattice) 
Coulomb potential and $H_{ij}(\vec{r})$ is an
angular potential. We regularize these potentials with a hard cut off : using 
$\theta (|\vec{r}|-a_{0})=1$ if $|\vec{r}|>a_{0}$ and $0$ otherwise, 
they are defined  as 
\begin{align}
\label{reg-potentials}
\tilde{G}(\vec{r}) & = \left(
\ln \left(\frac{|\vec{r}| }{a_{0}} \right) +c \right) \theta (|\vec{r}|-a_{0})
\quad ;\quad 
G(\vec{r})+i \Phi (\vec{r}) =  \ln \left(\frac{z }{a_{0}} \right) 
\theta (|\vec{r}|-a_{0})
\\
H_{ij}(\vec{r}) & = \left(\frac{r_{i}r_{j}}{r^{2}} - \frac{1}{2} \delta_{ij} \right)
\theta (|\vec{r}|-a_{0})
\end{align}
 where $z=r_{x} + i r_{y}$, $c$ is an arbitrary constant, and we have defined for later 
convenience the logarithmic potential $G(\vec{r})$. 
 In the presence of dislocations, the displacement field
splits into the above component $\vec{u}_{d}(\vec{r}) $ induced by the
dislocations themselves, and an independant smooth phonons part 
$\vec{u}_{ph}(\vec{r})$ : 
$\vec{u}(\vec{r})=\vec{u}_{ph}(\vec{r})+ \vec{u}_{d}(\vec{r}) $. 
 Without any perturbation, the usual melting transition is studied by
performing explicitly the integral over the smooth phonons field in
the partition function. One is left with the partitition function of
Coulomb gas with vector charges $\vec{b}_{\alpha}$, whose scaling
behavior describes the KTHNY melting transition. However, with
translation symmetry breaking perturbations, this usual (magnetic)
Coulomb gas must be extended to a electromagnetic gas, as explained
below.

\subsection{Breaking the translation symmetry}\label{sec:breaking}

 In this paper, we will consider a two dimensional crystal coupled to
a substrate modeled by a potential $V(\vec{r})$ coupling directly to
the density $\rho (\vec{r})$ 
of the lattice. This coupling adds to the elastic Hamiltonian
(\ref{eq:H-elastic}) an  energy 
\begin{equation}\label{eq:couplingV}
H_{V} = \int_{\vec{r}} \rho (\vec{r})  V (\vec{r} )
\end{equation}
 which explicitly depends on $\vec{u}$ instead of $u_{ij}$, reflecting
the breaking of the translation symmetry. 
 This symmetry breaking corresponds to the
situation where the density $\rho (\vec{r})$ and the potential $ V
(\vec{r} )$ have some harmonics in common corresponding to a reciprocal
lattice vector $\vec{G}$. In the following, we will consider either
the case of a periodic potential {\it commensurate} with the lattice,
or a random pinning potential. In both cases, we can consider that 
$\int_{\vec{r}} V(\vec{r})=0$. We decompose the lattice
density as 
\begin{equation}
\rho (\vec{r}) = \rho_{0} \left(
1-\partial_{i}u_{i}(\vec{r}) 
+\sum_{\vec{G}\neq 0 }e^{i\vec{G}.(\vec{r}-\vec{u}(\vec{r}))}
\right)
+ h.o.t.
\end{equation}
 where the $\vec{G}$ are reciprocal lattice vectors. 
 Similarly, the coupling (\ref{eq:couplingV}) reads : 
\begin{equation}\label{eq:couplingV-decomp}
\int_{\vec{r}} \rho (\vec{r})  V (\vec{r} ) = 
-\int_{\vec{r}}\left(\rho_{0}V(\vec{r})\right)
\partial_{i}u_{i}(\vec{r})
+ \frac{1}{2}\int_{\vec{r}}\sum_{\vec{G} \neq 0} \left(
V_{\vec{G} }e^{-i\vec{G}.\vec{u}(\vec{r})}
+V_{-\vec{G} }e^{i\vec{G}.\vec{u}(\vec{r})}
 \right)
\end{equation}
 where we have defined 
$V_{\vec{G}}=\rho_{0}V(\vec{r})e^{i\vec{G}.\vec{r}}$. Upon coarse
graining (or in an effective long wavelength hamiltonian), only the 
reciprocal lattice vectors common to $V(\vec{r})$ and $\rho(\vec{r}) $
will survive. In the above equation, the
primed sum is on these common 
reciprocal lattice vectors corresponding to a non
vanishing $V_{\vec{G}}$, which exists in the cases considered. 
 In the following, we will restrict ourselves only to these vectors in common 
$\vec{G}$ of minimum length. They correspond to the most
relevant perturbations near the pure melting transition.

\subsubsection{Periodic commensurate substrate}\label{sec:def-periodic}

In the case of a periodic and commensurate substrate,
we can use the symmetry $V_{\vec{G} }=V_{-\vec{G}}^{*}$ to 
rewrite  the  second term of (\ref{eq:couplingV-decomp}) as 
\begin{equation}\label{eq:couplingCos}
 \frac{H_{V}}{T}  = 
\int_{\vec{r}}\sum_{\vec{G}}
2 \frac{|V_{\vec{G}}|}{2T} \cos \left(\vec{G}.\vec{u}(\vec{r}) \right)
\end{equation}
As previously mentionned, we will restrict ourselves to the three
reciprocal lattice vectors $\vec{G}_{\alpha=1,2,3}$ of minimal length
$|\vec{G}_{1}|$ arising in this sum. 
 We will also use below the unit vectors 
$\hat{G}_{\alpha=1,2,3}=\vec{G}_{\alpha}/|\vec{G}_{\alpha}|$. 
In addition, 
a periodic substrate modifies the elastic part
of the energy by generating a new
term coupling the orientations of the lattice to the
substrate. Defining the local orientation 
$\theta(\vec{r})=\frac{1}{2}(\partial_{x}u_{y}-\partial_{y}u_{x}) $,
this new term can be written as 
\begin{equation}
\delta H_{0} = \frac{\gamma}{2}\int_{\vec{r} } \theta^{2}(\vec{r}) 
\end{equation}
where $\gamma$ is a new elastic constant. This term is non
zero even in the floating solid phase where the direct coupling
(\ref{eq:couplingCos}) is irrelevant, and must thus be included. 

Finally, we focus on the case of weak perturbations : to first
order in $V_{\vec{G}}$, we can expand the cosine coupling into 
\begin{align}\nonumber
\exp \left(2 \frac{|V_{\vec{G}_{\alpha}}|}{2T} 
\sum_{\alpha =1,2,3}
\cos \left(\vec{G}_{\alpha}.\vec{u}(\vec{r}) \right) \right)
& \simeq 
1+ \frac{|V_{\vec{G}_{\alpha}}|}{2T}
\sum_{\vec{m} (\vec{r})=\pm \hat{G}_{1},\pm \hat{G}_{2},\pm \hat{G}_{3}} 
e^{i |\vec{G}_{1}| \vec{m}(\vec{r}).\vec{u}(\vec{r})}\\
\label{eq:VillainPeriodic}
&= 
\sum_{\vec{m} (\vec{r})=0,\pm \hat{G}_{i=1,2,3}}
\left(\frac{|V_{\hat{G}_{\alpha}}|}{2T} 
\right)^{\vec{m}(\vec{r}).\vec{m}(\vec{r})}
e^{i  |\vec{G}_{1}| \vec{m}(\vec{r}).\vec{u}(\vec{r})}
\end{align}
 Defining a fugacity $Y[0,\vec{m}]$ for the formal charges
$\vec{m}(\vec{r})$ as  
\begin{equation}
Y[0,\vec{m}]= y^{\vec{m}.\vec{m}} \textrm{ with }
y=\frac{|V_{\vec{G}_{1}}|}{2T} 
\end{equation}
 we rewrite the partition function of the perturbed lattice as
 \begin{align}\label{eq:Zelecmag-0}
Z=&\int d[\vec{u}_{ph}(\vec{r})] \int d[\vec{u}_{d}(\vec{r})]
\left( \prod_{\vec{r}}\sum_{\vec{m}(\vec{r})} Y[0,\vec{m}(\vec{r})]  
\right)
 \\ \nonumber 
&
\exp \left(
 -\frac{1}{2 T}\int_{\vec{r}} 
\left[
u^{(d)}_{ij}(\vec{r})C_{ijkl}u^{(d)}_{kl}(\vec{r})
+u^{(ph)}_{ij}(\vec{r})C_{ijkl}u^{(ph)}_{kl}(\vec{r})
+u^{(d)}_{ij}(\vec{r})C_{ijkl}u^{(ph)}_{kl}(\vec{r})
 \right]
\right) \\ \nonumber 
& \exp \left( - \frac{1}{2 T} \int_{\vec{r}} \gamma \theta^2 \right)
\exp \left( i |\vec{G}_{1}|\int_{\vec{r}}\vec{m}(\vec{r}).
\left(\vec{u}^{(d)}+\vec{u}^{(ph)}   \right)(\vec{r})  
 \right)
\end{align}
Plugging the expression (\ref{eq:u_d}) for $\vec{u}^{(d)}$, and
integrating over the gaussian displacement field $\vec{u}^{(ph)}$, we
obtain three contributions to the remaining action: 
\begin{equation}\label{eq:Zelecmag}
Z = \sum_{\{\vec{b}(\vec{r}) \}} 
\left( \prod_{\vec{r}}\sum_{\vec{m}(\vec{r})} Y[0,\vec{m}(\vec{r})]  
\right)
\exp \left( S[\vec{b}/\vec{b}]+S[\vec{b}/\vec{m}]+ S[\vec{m}/\vec{m}]
\right)
\end{equation}
The dislocation interaction is given by the usual form extended to 
include the $\gamma$ coupling (see appendix
\ref{app:dislocations}) : 
\begin{align}
S[\vec{b}/\vec{b}] & = 
 -\frac{1}{2T}\int_{\vec{r}} 
u^{(d)}_{ij} (\vec{r})C_{ijkl}u^{(d)}_{kl}(\vec{r})\\
&= - \frac{a_{0}^{2}}{2T} 
 \int_{\vec{q} } b_{i} (\vec{q} )b_{j} (-\vec{q} ) \frac{1}{q^2} 
\left(
\frac{4 c_{66}\gamma }{c_{66}+\gamma} P_{ij}^{L}(\vec{q}) 
+ \frac{4c_{66} (c_{11}-c_{66})}{c_{11}} P_{ij}^{T}(\vec{q}) 
 \right)
\\\label{eq:Sbb}
& = 
\frac{1}{2}\sum_{\alpha \neq \beta} \left(
K_{1} \vec{b}_{\alpha}.\vec{b}_{\beta} 
G(\vec{r}_{\alpha}-\vec{r}_{\beta})    
-K_{2} b_{\alpha,i}b_{\beta,j}
H_{ij}(\vec{r}_{\alpha}-\vec{r}_{\beta}) \right)
-\frac{E_{c}}{T}\sum_{\alpha } \vec{b}_{\alpha}.\vec{b}_{\alpha}  
\end{align}
 where the inverse Fourier transform of  
\begin{equation}
f_{ij}(\vec{q}) = q^{-2}\left(A P_{ij}^{L}+ B P_{ij}^{T} \right)  
\end{equation}
 was determined as 
\begin{equation}
f_{ij}(\vec{r})  = \int_q (1-e^{i \vec q \cdot \vec r}) f_{ij}(\vec{q}) =
\delta_{ij}\frac{A+B}{4\pi}\left(\ln \frac{r}{a}+cte \right) 
+\frac{A-B}{4\pi} H_{ij}(\vec{r}) 
\end{equation}
providing the following expressions for the coupling constants 
\begin{equation}
K_{1/2} = \frac{a_{0}^{2}}{\pi T}\left( \frac{c_{66}(c_{11}-c_{66})}{c_{11}} 
\pm \frac{c_{66}\gamma }{c_{66}+\gamma }\right)
=  \frac{a_{0}^{2}}{\pi T}\left(\frac{\mu (\mu +\lambda)}{2\mu +\lambda}
\pm \frac{\mu \gamma}{\mu +\gamma }\right)
\end{equation}
Note that the dislocation core energy $E_{c}$ in (\ref{eq:Sbb}) arises
from the standard continuum approximation (\ref{reg-potentials}) 
of the lattice Coulomb interaction $\tilde{G}(r)$, and the use of 
the neutrality condition $\int_{\vec r} \vec b(\vec r) = 0$. From now on, the core energy $E_{c}$ 
will be
incorporated in a fugacity for the $\vec{b}$ charges: 
\begin{equation}
Y[\vec{b},\vec{0}]=
\tilde{y}^{\vec{b}.\vec{b}} 
\quad \textrm{with}\quad 
\tilde{y} = e^{-\frac{E_{c}}{T}}  . 
\end{equation}
 The interaction between the $\vec{m}$ charges follows from the
gaussian integration over $\vec{u}^{(ph)}$ : 
\begin{align}
S[\vec{m}/\vec{m}] &= 
-\frac{|\vec{G}_{1}|^{2}}{2}\int_{\vec{q}} 
m_{i}(\vec{q}) \Phi_{ij}^{-1} m_{j}(\vec{q})  \\
&= -\frac{1}{2}\int_{\vec{q}} 
m_{i}(\vec{q})
\left( 
 \frac{|\vec{G}_{1}|^{2} T}{c_{11}q^{2}}P_{ij}^{L} 
+\frac{|\vec{G}_{1}|^{2} T}{(c_{66}+\gamma)q^{2}}P_{ij}^{T} \right)
m_{j}(\vec{q}) \\
& = 
\frac{1}{2}\sum_{\alpha \neq \beta} 
\left(
K_{3} \vec{m}_{\alpha}.\vec{m}_{\beta} 
G(\vec{r}_{\alpha}-\vec{r}_{\beta})    
-K_{4} m_{\alpha,i}m_{\beta,j}
H_{ij}(\vec{r}_{\alpha}-\vec{r}_{\beta}) 
\right) 
\nonumber
\\ 
& \hspace{1cm} 
-\frac{\tilde{E}_{c}}{T}\sum_{\alpha } \vec{m}_{\alpha}.\vec{m}_{\alpha}  
\end{align}
 with the coupling constants 
\begin{equation}
K_{3/4} = 
\frac{T|\vec{G}_{1}|^{2} }{4\pi}\left(
\frac{1}{c_{66} +\gamma} \pm \frac{1}{c_{11}}
 \right)
=
\frac{T|\vec{G}_{1}|^{2} }{4\pi}\left(
\frac{1}{\mu +\gamma} \pm \frac{1}{2\mu +\lambda}
 \right)
\end{equation}
 The core energy $\tilde{E}_{c}$ will incorporated from now on into
the bare fugacity $Y[\vec{0},\vec{m}]$. 
 Finally the cross coupling comes from the last term in\footnote{
Note that, as a consequence of the hard-core regularization of the potentials 
(\ref{reg-potentials}), one can use indifferently  a sum over distinct 
charges ($\sum_{\alpha\neq \beta}$) or not ($\sum_{\alpha,\beta}$) in 
the expression below.} 
(\ref{eq:Zelecmag-0}) : 
\begin{align}
S[\vec{b}/\vec{m}] &= 
 \frac{i a_{0}|\vec{G}_{1}| }{2\pi} \sum_{\alpha,\beta} 
m_{i}(\vec{r}_{\alpha})   \mathcal{G}_{ij}
(\vec{r}_{\alpha }-\vec{r}_{\beta})
b_{j}(\vec{r_{\beta}})    \\ \nonumber 
&= i \sum_{\alpha,\beta} 
m_{i}(\vec{r}_{\alpha})
\biggl( \delta_{ij} \frac{a_{0}|\vec{G}_{1}|}{2\pi}\Phi(\vec{r}_{\alpha }-\vec{r}_{\beta}) 
+ K_{5}
 \epsilon_{ij} G(\vec{r}_{\alpha }-\vec{r}_{\beta}) 
 \\
&
+ K_{6} \epsilon_{jk} 
 H_{ik}(\vec{r}_{\alpha }-\vec{r}_{\beta}) \biggr)
b_{j}(\vec{r_{\beta}}) 
\end{align}
 with 
\begin{equation}
K_{5} = \frac{a_{0}|\vec{G}_{1}|}{2\pi} (\frac{c_{66}}{ c_{11}} - \frac{\gamma}{\gamma + c_{66}} )
\quad ;\quad 
K_{6} =  \frac{a_{0}|\vec{G}_{1}|}{2\pi} ( \frac{ c_{11} - c_{66}}{c_{11}}  
- \frac{\gamma}{\gamma + c_{66}} )
\end{equation}
 Defining the potential 
\begin{equation}\label{def-potential}
V_{ij}(K_{1},K_{2},\vec{r}) = 
K_{1} \delta_{ij} G(\vec{r})  
-K_{2} H_{ij}(\vec{r}) 
\end{equation}
 we can rewrite the above partition function as that of a Coulomb gas
with both electric and magnetic vector charges :  
\begin{equation}
Z = \sum_{  \{ \vec r_\alpha , \vec{b}(\vec{r}_\alpha),\vec{m}(\vec{r}_\alpha)\}} 
\prod_{\alpha} Y[\vec{b}_{\alpha},\vec{m}_{\alpha}]  
\exp S[\vec{b}(\vec{r}_\alpha),\vec{m}(\vec{r}_\alpha)]
\end{equation}
 with the action 
\begin{align}
S[\vec{b}(\vec{r}_\alpha),\vec{m}(\vec{r}_\alpha)] &= 
\frac{1}{2} \sum_{\alpha\neq \beta}
b_{\alpha,i} V_{ij}(K_{1},K_{2},\vec{r}_{\alpha}-\vec{r}_{\beta})
 b_{\beta,j}   
 \nonumber \\
 &
+\frac{1}{2}\sum_{\alpha\neq \beta}
m_{\alpha,i} V_{ij}(K_{3},K_{4},\vec{r}_{\alpha}-\vec{r}_{\beta})
m_{\beta,j}   
 \nonumber \\
&  
+ i \sum_{\alpha\neq \beta}
m_{\alpha,i} \left(
\delta_{ij}\frac{a_{0}|\vec{G}_{1}|}{2\pi}\Phi(\vec{r}_{\alpha}-\vec{r}_{\beta}) 
+ V_{ik}(K_{5},K_{6})\epsilon_{kj}
\right)b_{\beta,j} \label{cgdescription}
\end{align}

\subsubsection{Substrate Disorder}\label{sec:def-disorder}

 In the case of a substrate disorder, the potential $V(\vec{r})$ which
couples to the local density of atoms of the crystal is random : its distribution
will be taken as gaussian, with variance 
\begin{equation}
 \overline{V (\vec{r} )V (\vec{r}' )}=h (\vec{r}-\vec{r}')
\end{equation}
where $h (\vec{r}-\vec{r}')$ is a short range correlator
and here and below $\overline{...}$ denotes an average over the disorder $V$. The two first
contributions from the Fourier decomposition
(\ref{eq:couplingV-decomp})  are 

\begin{equation}\label{eq:H-pinning}
\frac{H_{V}[\vec u]}{T}
=  \int_{\vec{r}} \left(
\frac{1}{T}\sigma_{ij}u_{ij} 
+ 2 \sqrt{y_{m}} \sum_{\nu=1,2,3} 
\cos \left( \vec{G}_{\nu} .\vec{u} (\vec{r}) 
   + \phi_{\nu} (\vec{r}) \right) \right)
\end{equation}
where $\sigma_{ij}$ is a random stress field, arising from the long 
wavelength part of the disorder potential $V(\vec{r})$.  It induces local 
random compression/dilation and shear stress.  Its correlator is 
parametrized as 
\begin{equation} \label{eq:corel-stress}
\overline{ \sigma_{ij}(\vec{r}) \sigma_{kl}(\vec{r}')} = 
\delta(\vec{r}-\vec{r}') \left[ (\Delta_{11}- 2 \Delta_{66}) 
\delta_{ij} \delta_{kl}+ \Delta_{66} ( 
\delta_{ik}\delta_{jl}+\delta_{il}\delta_{jk}) \right]
\end{equation}
whose bare values, derived from (\ref{eq:couplingV-decomp})
 are:
\begin{equation}
\Delta_{11}=\rho_0^2 h_{\vec{K} =\vec{0} }
\quad ;\quad 
\Delta_{66}=0
\quad ;\quad 
y_m=\rho_0^2 h_{\vec{K} =\vec{G}_{1} }/T^2
\end{equation}
where $\rho_0$ is the mean density. The second part 
of the disorder comes from the first harmonic of $V(\vec{r})$ with almost 
the same periodicity as the lattice, i.e. it is proportional to $\sqrt{y_{m}}$ the amplitude
of the $\vec{q}\simeq \vec{G}_{1}$ component of $V(\vec{r})$ occuring in
(\ref{eq:couplingV-decomp}). Since it is not invariant under a uniform
shift of $\vec u$ it is usually called the {\it pinning disorder}.  The 
random phase field in (\ref{eq:corel-stress}) is uniformly distributed 
over $[0,2\pi]$ and satisfies 
\begin{equation}
\overline{ \langle e^{i(\phi_{\nu}({\bf r})-\phi_{\nu'}(\vec{r}'))} \rangle } 
= \delta_{\nu,\nu'}  \delta^{2}(\vec{r}-\vec{r}') . 
\end{equation}
The $\vec{G}_{\nu}$ are the first 
reciprocal lattice vectors (of modulus $G_{1}^{2}= 16 \pi^{2} /3 
a_{0}^{2}$ ). 

The average over the disorder fields $\phi_{\nu}(\vec{r})$ and 
$ \sigma_{ij}(\vec{r})$ is performed using the replica trick introducing the
replicated field 
$\vec{u}^a(\vec{r})$, $a=1,...n$, and the corresponding replicated Burgers charge
$\vec{b}^a(\vec{r})$. One defines:
\begin{align} \label{zdis}
Z= \overline{Z_V^n} = \overline{\prod_{a=1}^n \int d[\vec u_a]
\exp\left(\frac{H_0[\vec u_a] + H_{V}[\vec u_a]}{T}\right)}
\end{align}
and consider the limit $n=0$. We focus on the case of  weak pinning disorder $y_{m}$, and 
expand the exponential of the 
 cosine coupling in (\ref{eq:H-pinning}) as in (\ref{eq:VillainPeriodic}) and perform the
disorder average in (\ref{zdis}):
 \begin{align}
&
\overline{
\exp \left( - 2 \sqrt{y_{m}} \sum_{\nu=1,2,3}\sum_{a=1}^{n} \cos \left(
\vec{G}_{\nu}.\vec{u}^a(\vec{r}) + \phi_{\nu} (\vec{r}) \right) \right) } 
\\
& = 
1+ y_{m} \sum_{a,b=1}^{n} \sum_{\nu=1,2,3}
e^{i\vec{G}_{\nu}.(\vec{u}^{a}(\vec{r})-\vec{u}^{b}(\vec{r}))}
+\mathcal{O}(y_{m}^{2})
\\
& = n y_{m} + 
\sum_{\vec{m}^{a} (\vec{r})}
Y[0,\vec{m}^{a}] 
e^{-i |\vec{G}_{1}| \sum_a \vec{m}^{a} \cdot \vec{u}^{a} }
 \end{align} 
The  replicated $\vec{m}$ charges have initially two opposite non zero components: 
\begin{equation}
\vec{m}^{a} = \hat{G}_{\nu} \left( \delta_{a,b_{1}} - \delta_{a,b_{2}} \right) 
\textrm{ with } 
b_{1} \neq b_{2}, 1 \leq b_{1},b_{2} \leq n, \nu =1,2,3 
\end{equation}
 However, under the fusion process of the renormalization procedure, we will have to consider 
 charges obtained as the sum of these initial charges. These general charges will be characterized by the 
 property $\sum_{a} \vec{m}^{a} = \vec{0}$. Their bare fugacity, introduced in the above formula, reads 
\begin{equation}
Y[0,\vec{m}^{a}] = \sqrt{y_{m}}^{\sum_{a}\vec{m}^{a}.\vec{m}^{a}}
\end{equation}

To introduce dislocations one can now follow the same steps as in Section 
\ref{sec:def-periodic} splitting $\vec u_a = \vec u^{(ph)}_a + 
\vec u_a^{(d)}$. The average over the random stress tensor (\ref{eq:corel-stress}) leads to
the replicated elastic matrices 
 \begin{equation}
 c_{11}^{ab} = c_{11} \delta^{ab} - \Delta_{11}
 \quad ; \quad 
  c_{66}^{ab} = c_{66} \delta^{ab} - \Delta_{66}
   \quad ; \quad 
  \gamma^{ab} = \gamma \delta^{ab} - \Delta_{\gamma} \label{elastmat}
 \end{equation}
 Hence, by the same technique as in the case of the commensurate
regular substrate, we obtain a Coulomb gas description (\ref{cgdescription}) of the
random model, albeit with coupling constant $K_{1,...6}$ which are
now replica matrices involving products and inverses of the replica
elastic matrices (\ref{elastmat})
\begin{subequations}
\label{replicaCG}
\begin{eqnarray}
&& K_{1/2} = \frac{a_{0}^{2}}{\pi T} \left( c_{66} (c_{11}-c_{66}) c_{11}^{-1}  
\pm c_{66} \gamma (c_{66} + \gamma)^{-1} \right) \\
&& K_{3/4} = 
\frac{T|\vec{G}_{1}|^{2} }{4\pi}\left(
(c_{66} +\gamma)^{-1} \pm c_{11}^{-1} \right) \\
&& K_{5} = \frac{a_{0}|\vec{G}_{1}|}{2\pi} (c_{66} c_{11}^{-1} - \gamma (\gamma + c_{66})^{-1} ) \\
&& K_{6} =  \frac{a_{0}|\vec{G}_{1}|}{2\pi} ( (c_{11} - c_{66}) c_{11}^{-1} 
- \gamma (\gamma + c_{66})^{-1} )
\end{eqnarray}
\end{subequations}
and thus contain information both about elastic constants
and longwavelength disorder. The only other modification is
the nature of the $\vec{m}$ charges, detailed above.

\subsection{Electromagnetic Coulomb gas with vector charges}\label{sec:VECG}

\subsubsection{Definition}\label{sec:defVECG}

 To study the scaling behaviour of the two above models with and  without disorder, it appears
necessary to consider a general electromagnetic Coulomb gas with
vector charges. In full generality, we will consider replicated
charges $\vec{b}^{a},\vec{m}^{a}$ of $n$ components.  Each component
of the Burgers charges $\vec{b}^{a}$ lies on the direct lattice,
while components of the $\vec{m}^{a}$ charges are reciprocal lattice
vectors. Any additional condition on the allowed charges, specific to
the model considered, will be detailed at a later stage of the
study. Our derivation of the renormalization equations will stick to
the most general model. The partition function of this Coulomb gas
 is defined by
\begin{equation} \label{eq:Z-VECG-lattice}
Z = \sum_{\{\vec{r}_{\alpha}, \vec{b}^{a}_{\alpha}(\vec{r}_{\alpha}),
\vec{m}^{a}_{\alpha}(\vec{r}_{\alpha})\}} 
\prod_{\alpha}Y[\vec{\bf b}_{\alpha},\vec{\bf m}_{\alpha}]  
\exp S[\vec{b}^{a}_{\alpha}(\vec{r}_{\alpha}),\vec{m}^{a}_{\alpha}(\vec{r}_{\alpha})]
\end{equation}
 where the sum counts each configuration of indistinguishable charges only once. 
These configurations  
 correspond to electromagnetic charges $\vec{b}^{a}_{\alpha},\vec{m}^{a}_{\alpha}$, 
labelled by the index  
 $\alpha$, both located in $\vec{r}_{\alpha}$ which belongs either to a lattice 
(lattice Coulomb gas) or to the continuum plane with a hard core constraint 
(see eq. (\ref{reg-potentials})). These configurations satisfy a neutrality condition : 
 \begin{equation}
 \sum_{\alpha} \vec{b}^{a}_{\alpha} =  \sum_{\alpha} \vec{m}^{a}_{\alpha} = \vec{0} 
 \textrm{  for each } a=1,...,n
 \end{equation}
 The action of this Coulomb gas reads 
\begin{align}
S[\vec{b}_{\alpha}(\vec{r}_{\alpha}),\vec{m}_{\alpha}(\vec{r}_{\alpha})] &= 
\frac{1}{2} \sum_{\alpha\neq \beta}
b_{\alpha,i}^{a} V_{ij}(K_{1}^{ab},K_{2}^{ab},\vec{r}_{\alpha}-\vec{r}_{\beta})
 b_{\beta,j}^{b}   
 \nonumber \\
 &
+\frac{1}{2}\sum_{\alpha\neq \beta}
m_{\alpha,i}^{a} V_{ij}(K_{3}^{ab},K_{4}^{ab},\vec{r}_{\alpha}-\vec{r}_{\beta})
m_{\beta,j}^{b} 
 \nonumber \\
&  
+ i \sum_{\alpha\neq \beta}
m_{\alpha,i}^{a} \left(
\delta_{ij}\delta^{ab}
\frac{\lambda_{\Phi}}{2\pi}\Phi(\vec{r}_{\alpha}-\vec{r}_{\beta}) 
+ V_{ik}(K_{5}^{ab},K_{6}^{ab})\epsilon_{kj}
\right)b_{\beta,j}^{b} 
\label{eq:S-VECG-lattice}
\end{align}
 where the interaction potentials $V_{ij}$  has been defined in (\ref{def-potential}), and 
the coupling matrices $K_{i}$ in section 
(\ref{sec:def-periodic}) for the commensurate potential, and in 
(\ref{replicaCG}) for the pinning random potential. We also define 
the geometrical factor 
\begin{equation}
\lambda_{\Phi} = a_{0}|\vec{G}_{1}| . 
\end{equation}
 where $\lambda_{\Phi}= 4\pi /\sqrt{3}$ for the triangular lattice, 
and $\lambda_{\Phi}= 2\pi$ for the square lattice. 
 Defining charge densities as 
 \begin{equation}
 \vec{b}^{a}(\vec{r}) = \sum_{\alpha}  \vec{b}^{a}_{\alpha}\delta(\vec{r}-\vec{r}_{\alpha})
 \quad ; \quad 
 \vec{m}^{a}(\vec{r}) = \sum_{\alpha}  \vec{m}^{a}_{\alpha}\delta(\vec{r}-\vec{r}_{\alpha})
 \end{equation} 
we can express this partition function as 
\begin{equation} \label{eq:Z-VECG}
Z = \sum_{\{\vec{b}^{a}(\vec{r}),\vec{m}^{a}(\vec{r})\}} 
\exp \left(\int \frac{d^{2}\vec{r}}{a_0^2}  \ln Y[\vec{b}^{a}(\vec{r}),\vec{m}^{a}(\vec{r})]  \right)
\exp S[\vec{b}^{a}(\vec{r}),\vec{m}^{a}(\vec{r})]
\end{equation}
with\footnote{Note that 
the angle $\Phi$ being defined up to a constant, the model is defined for configurations 
satisfying $\sum_{\alpha} \sum_a \vec{b}^{a}_{\alpha}.\vec{m}^{a}_{\alpha} = 0$. This condition is 
satisfied in a bare model consisting of a collection of purely electric 
($\vec{\bf m}=\vec{0}$) and purely magnetic ($\vec{\bf b}_{\alpha}=\vec{0}$) charges. 
Without this condition, a change of definition of the angle $\Phi\to \Phi + \theta_{0}$ is accompanied 
by a redefinition of the fugacities for composites charges : 
$Y[b,m]\to Y[b,m] \exp[-i \theta_{0} b.m].$
} 
\begin{align}
S[\vec{b}(\vec{r}),\vec{m}(\vec{r})] &=
\frac{1}{2}
b_{i}^{a} * V_{ij}(K_{1}^{ab},K_{2}^{ab}) * b_{j}^{b}   
+\frac{1}{2}
m_{i}^{a} * V_{ij}(K_{3}^{ab},K_{4}^{ab}) * m_{j}^{b} 
 \nonumber \\
&  
+ i 
m_{i}^{a} * \left(\delta_{ij}\delta^{ab}\frac{\lambda_{\Phi}}{2\pi}\Phi 
+ V_{ik}(K_{5}^{ab},K_{6}^{ab})\epsilon_{kj}\right) * b_{j}^{b} 
 \label{eq:S-VECG}
\end{align}

 This is a vector generalisation of the 2D scalar electromagnetic 
 coulomb gas and of the electric vector coulomb gas which enter the 
 standard study of melting. 

\subsubsection{Electromagnetic duality}\label{duality}

In 2D coulomb gas, the Kramers-Wannier duality corresponds to the 
interchange of electric and magnetic charges : $\vec{b} \leftrightarrow \vec{m}$.  
In the usual scalar ECG, this corresponds 
to the interchange of strong and weak coupling regimes of the theory :  
$g \leftrightarrow 1/g$ where $g$ is the  
  coupling constant of the ECG. For the present general VECG, this duality 
transformation can be inferred by 
by writing explicitly the action (\ref{eq:S-VECG}) as 
\begin{align*}
S[\vec{b}(\vec{r}),\vec{m}(\vec{r})] &=
\frac{1}{2}\sum_{\alpha\neq \beta} \left[
 K_{1}^{ac}(\vec{b}^{a}_{\alpha}.\vec{b}^{c}_{\beta}) G(r_{\alpha\beta}) 
-  K_{2}^{ac} \left( (\vec{b}^{a}_{\alpha}.\hat{r}_{\alpha\beta})(\vec{b}^{c}_{\beta}.\hat{r}_{\alpha\beta})
- \frac12 (\vec{b}^{a}_{\alpha}.\vec{b}^{c}_{\beta}) \right) \right]\\
&+\frac{1}{2}\sum_{\alpha\neq \beta} \left[
 K_{3}^{ac}(\vec{m}^{a}_{\alpha}.\vec{m}^{c}_{\beta}) G(r_{\alpha\beta}) 
-  K_{4}^{ac} \left( (\vec{m}^{a}_{\alpha}.\hat{r}_{\alpha\beta})(\vec{m}^{c}_{\beta}.\hat{r}_{\alpha\beta})
- \frac12 (\vec{m}^{a}_{\alpha}.\vec{m}^{c}_{\beta}) \right) \right]\\
& + i \sum_{\alpha\neq \beta} \big[
(\vec{m}^{a}_{\alpha}.\vec{b}^{a}_{\beta}) \frac{\lambda_{\Phi}}{2\pi}\Phi (\vec{r}_{\alpha\beta})  \\
& + 
 K_{5}^{ac}  (\vec{m}^{a}_{\alpha}.(\vec{b}^{\perp})^{c}_{\beta})G(\vec{r}_{\alpha\beta}) 
-K_{6}^{ac}
 \left( (\vec{m}^{a}_{\alpha}.\hat{r}_{\alpha\beta}) ((\vec{b}^{\perp})^{c}_{\beta}.\hat{r}_{\alpha\beta})
-\frac12 (\vec{m}^{a}_{\alpha}.(\vec{b}^{\perp})^{c}_{\beta}) \right) \big]
\end{align*}
 with the convention $a^{\perp}_{i}=\epsilon_{ij} a_{j}$. 
Inspection of the above expression, and the relation\footnote{Note
also the useful relation 
$\hat{r}_i \hat{r}^{\perp}_j - \hat{r}_j \hat{r}^{\perp}_i =- \epsilon_{ij}$}
$\hat{r}_i \hat{r}_j + \hat{r}^{\perp}_i \hat{r}^{\perp}_j =\delta_{ij}$
 (or $H_{ij}(\hat{r}^{\perp})=-H_{ij}(\hat{r})$),  
shows that performing the simultaneous change:
\begin{equation}
(\vec{\bf b}_{\alpha},\vec{\bf m}_{\alpha}) \rightarrow 
(\vec{\bf b}'_{\alpha} = \vec{\bf m}_{\alpha}^{\perp},
\vec{\bf m}'_{\alpha}=\vec{\bf b}^{\perp}_{\alpha})
\end{equation}
 and 
\begin{eqnarray}
&& K_{1}\rightarrow K_{1}'=K_{3} \quad , \quad K_{3}\rightarrow K_{3}'=K_{1} \\
&& 
K_{2} \rightarrow K_{2}' = - K_{4} \quad , \quad K_{4}\rightarrow K_{4}'=- K_{2} \\
&& 
K_{5}\rightarrow K_{5}'=- K_{5} \quad , \quad K_{6} \rightarrow K_{6}'=- K_{6}
\end{eqnarray}
leaves the action unchanged. This is the duality transformation. 
Note that the symmetry by orientation change $B \rightarrow -B$ (or time 
 reversal) corresponds to $i \rightarrow -i$.  It affects only the 
$\vec{b}/\vec{m}$ interaction.

\section{Renormalization of the Coulomb gas}\label{sec:renormalisation}

 The renormalization  of this electromagnetic Coulomb gas goes along the lines of the Coulomb gas with 
 scalar charges (Nienhuis) : upon increasing the real space cut-off $a_{0}\rightarrow a_{0}e^{dl}$ 
 (corresponding to the size of 
 the charges), we have to consider three different processes : 
(i) the simple rescaling of the partition functions's integration measures 
 and the Coulomb interaction, 
(ii) the screening or annihilation of charges, corresponding to the modification 
 of the Coulomb interaction of distant charges by two opposite charges distant by less 
than the new cut-off
 $a_{0}e^{dl}$, 
and (iii) the fusion of charges when two non-opposite charges distant by less than the new cut-off
 have to be considered as a new single charge at the new scale. We will consider successively this three 
 processes.  

\subsection{Reparametrization}\label{sec:reparametrisation}

Simple rescaling of the cut-off $a_{0}\rightarrow a_{0}e^{dl}$ into the integration measure 
($d^2 \vec{r}/a_{0}^{2}$) and 
and  the Coulomb interaction (from the terms containing $\ln(r/a_{0})$)
results in  the eigenvalue
\begin{equation}
\label{eq:repara}
\partial_l Y[\vec{\bf b},\vec{\bf m}] = 
\left( 2-
\frac12
\left( 
\vec{b}^{a}.\vec{b}^{b}  K_{1}^{ab} + 
\vec{m}^{a}.\vec{m}^{b}  K_{3}^{ab}
+2 i m^{a}_{i} \epsilon_{ij} b_{j}^{b} K_{5}^{ab}
\right)\right)Y[\vec{\bf b},\vec{\bf m}]
\end{equation}

\subsection{Fusion of charges}\label{sec:fusion}
 
 We consider the situation where two charges $(\vec{\bf b}_{1},\vec{\bf m}_{1})$ and 
 $(\vec{\bf b}_{2},\vec{\bf m}_{2})$ located in $\vec{r}_{1}$ and $\vec{r}_{2}$ are distant by less than the rescaled cutoff : 
 $a_{0}< |\vec{\rho}| < a_{0}e^{dl}$
where we define $\vec{\rho}=\vec{r}_{1}-\vec{r}_{2}$.  The part $\tilde{S}_{12}$ 
 of the action (\ref{eq:S-VECG})  involving these two charges can be decomposed into their mutual 
 interaction and the interaction with the rest of the charge configuration 
 $\tilde{S}_{12}=S_{1,2}+\sum_{\alpha\neq 1,2}S_{1,2/\alpha}$ . 
 From now on, we will use the notation
 \begin{align}
& V_{(1),ij}^{ab} = V_{ij}(K_{1}^{ab}, K_{2}^{ab}) 
 \quad ; \quad 
  V_{(3),ij}^{ab} = V_{ij}(K_{3}^{ab}, K_{4}^{ab}) 
\\ 
\label{eq:G-V}
&\mathcal{G}_{ij}^{ab} 
 =  \delta_{ij}\delta^{ab}\frac{\lambda_{\Phi}}{2\pi} \Phi + \epsilon_{kj} V_{ik}( K_5^{ab}, K_6^{ab})
 \end{align}
  With this notation, the mutual interaction between charges $1$ and $2$ reads  
\begin{multline}
S_{1,2}(\vec{\rho}) =  
b_{1,i}^{a}  V_{(1),ij}^{ab}(\vec{\rho}) b_{2,j}^{b} 
+ m_{1,i}^{a}  V_{(3),ij}^{ab}(\vec{\rho}) m_{2,j}^{b} \\
+ i ~\left( 
m_{1,i}^{a} \mathcal{G}_{ij}^{ab}(\vec{\rho})  b_{2,j}^{b} 
+ m_{2,i}^{a} \mathcal{G}_{ij}^{ab}(-\vec{\rho})   b_{1,j}^{b} 
\right) \label{S-pq} 
\end{multline}
Similarly the interaction between this pair and another charge $\alpha$ is written as 
\begin{multline}
S_{1,2/\alpha}  = 
b_{1,i}^{a}  V_{(1),ij}^{ab}(\vec{r}_{1}-\vec{r}_{\alpha}) b_{\alpha,j}^{b} 
+ m_{1,i}^{a}  V_{(3),ij}^{ab}(\vec{r}_{1}-\vec{r}_{\alpha}) m_{\alpha,j}^{b} 
\\
+ i ~\left( 
m_{1,i}^{a} \mathcal{G}_{ij}^{ab}(\vec{r}_{1}-\vec{r}_{\alpha}) b_{\alpha,j}^{b} 
+ m_{\alpha,i}^{a} \mathcal{G}_{ij}^{ab}(\vec{r}_{\alpha}-\vec{r}_{1}) b_{1,j}^{b} 
\right)
+
(1 \leftrightarrow 2)
\label{eq:S-alpha}
\end{multline}
The part of the partition function involving the 
 two charges $(\vec{\bf b}_{1},\vec{\bf m}_{1})$ and 
 $(\vec{\bf b}_{2},\vec{\bf m}_{2})$ can be written as\footnote{Note that the multiple integral should 
 be restricted to the domain $|\vec{r}_{\alpha}-\vec{r}_{\beta}|\geq a_{0}$}
 \begin{multline}
Z_{1,2} = 
\sum_{(\vec{\bf b}_{1/2},\vec{\bf m}_{1/2})\in \{\vec{\bf b}_{\alpha},\vec{\bf m}_{\alpha}\}}
\left( \prod_{\alpha} \int \frac{d^{2}\vec{r}_{\alpha}}{a_{0}^{2}} \right)
\\
\prod_{\alpha \neq 1,2} Y[\vec{\bf b}_{\alpha},\vec{\bf m}_{\alpha}]
 Y[\vec{\bf b}_{1},\vec{\bf m}_{1}]Y[\vec{\bf b}_{2},\vec{\bf m}_{2}] 
e^{S_{1,2}+\sum_{\alpha\neq 1,2}S_{1,2/\alpha}}
\label{eq:Z-pq}
 \end{multline}
 We are interested in the correction of order $dl$ coming from this partial partition function. 
To proceed, two cases must be distinguished : either the total charge in non zero, or 
$\vec{\bf b}_{1}+\vec{\bf b}_{2}=\vec{\bf m}_{1}+\vec{\bf m}_{2}=\vec{0}$.  
The first case corresponds to the fusion of charges considered below, 
and the second to the annihilation of charges (or Debye screening of the 
interactions), which will be considered in the next section. 

 In the first case we have 
$\vec{\bf b}_{1}+\vec{\bf b}_{2}\neq 0$ or/and $\vec{\bf m}_{1}+\vec{\bf m}_{2}\neq 0$.  
This gives after coarse graining a non zero effective charge located in 
$\vec{R}=(\vec{r}_{1}+\vec{r}_{2})/2$. 
 To proceed, we assume a low density for the Coulomb gas, which amounts to 
consider that all interdistances $\vec{r}_{\alpha}-\vec{r}_{\beta}$ 
between the remaining charges are much larger than $a_{0}$. This allows to 
perform a gradient expansion of the integrand 
$\exp \left( S_{1,2}+\sum_{\alpha\neq 1,2}S_{\alpha}\right)$. The first non-vanishing 
term of this expansion is simply the term of order $0$ for the fusion of charges.   
 To this order, the correction  (\ref{eq:Z-pq}) simply  reads 
\begin{multline}\label{eq:Z-12-fusion}
Z_{1,2} = dl
\sum_{(\vec{\bf b}_{1/2},\vec{\bf m}_{1/2})\in \{\vec{\bf b}_{\alpha},\vec{\bf m}_{\alpha}\} 
\atop
(\vec{\bf b}_{1},\vec{\bf m}_{1})+(\vec{\bf b}_{2},\vec{\bf m}_{2})\neq (\vec{0},\vec{0})} 
\left( \prod_{\alpha\neq 1,2} \int \frac{d^{2}\vec{r}_{\alpha}}{a_{0}^{2}}
Y[\vec{\bf b}_{\alpha},\vec{\bf m}_{\alpha}] \right)
 \int \frac{d^{2}\vec{R}}{a^{2}} 
\\
Y[\vec{\bf b}_{1},\vec{\bf m}_{1}]Y[\vec{\bf b}_{2},\vec{\bf m}_{2}]
\left( \int d\hat{\rho} e^{S_{1,2}} \right)
e^{\sum_{\alpha\neq 1,2}S_{1,2/\alpha}}
+\mathcal{O}(dl^{2})
\end{multline}
 where we have used the notation $\int d\hat{\rho}$ for the integral on the unit 
circle $\int_{0}^{2 \pi}d\theta_{\vec{\rho}}$. 
 The term (\ref{eq:Z-12-fusion}) 
will correct the partition function over the same final configuration of charges, including the 
 new effective charge in $\vec{R}$. To order $0$ in the gradient expansion, 
 $\sum_{\alpha\neq 1,2}S_{1,2/\alpha}$ provides exactly the correct interaction between the new 
 charge and the rest of the configuration. Thus the above partition function can be absorbed into a 
 correction to the fugacity for non zero charges 
\begin{equation}
\partial_{l}Y[\vec{\bf b},\vec{\bf m}] =
\sum_{(\vec{\bf b}_{1},\vec{\bf m}_{1})+(\vec{\bf b}_{2},\vec{\bf m}_{2})= 
(\vec{\bf b},\vec{\bf m})}
A_{(\vec{\bf b}_{1},\vec{\bf m}_{1});(\vec{\bf b}_{2},\vec{\bf m}_{2})} 
Y[\vec{\bf b}_{1},\vec{\bf m}_{1}]Y[\vec{\bf b}_{2},\vec{\bf m}_{2}]
\label{RG-fusion}
\end{equation}
where the numerical factor 
\begin{equation}\label{A-pq}
A_{(\vec{\bf b}_{1},\vec{\bf m}_{1});(\vec{\bf b}_{2},\vec{\bf m}_{2})}= 
\int d\hat{\rho} 
\exp(S[(\vec{\bf b}_{1},\vec{\bf m}_{1});(\vec{\bf b}_{2},\vec{\bf m}_{2})])
\end{equation}
 with the action $S[(\vec{\bf b}_{1},\vec{\bf m}_{1});(\vec{\bf b}_{2},\vec{\bf m}_{2})]$ given 
by (\ref{S-pq}) with $\rho = a_{0}$ :  
\begin{align} \nonumber
S[(\vec{\bf b}_{1},\vec{\bf m}_{1});(\vec{\bf b}_{2},\vec{\bf m}_{2})] = &
- (\vec{\bf b}_{1})^{a}_{i} (\vec{\bf b}_{2})^{b}_{j} K_{2}^{ab} H_{ij} (\hat{\rho})
- (\vec{\bf m}_{1})^{a}_{i} (\vec{\bf m}_{2})^{b}_{j} K_{4}^{ab} H_{ij} (\hat{\rho})
\\ \nonumber
&+ i (\vec{\bf m}_{1})^{a}_{i} (\vec{\bf b}_{2})^{a}_{i}
\frac{\lambda_{\Phi}}{2\pi} \Phi(\hat{\rho})
 - i (\vec{\bf m}_{1})^{a}_{i} (\vec{\bf b}_{2})^{b}_{j} K_{6}^{ab}
    \epsilon_{kj} H_{ik}(\hat{\rho})
\\
&+ i (\vec{\bf m}_{2})^{a}_{i} (\vec{\bf b}_{1})^{a}_{i}
\frac{\lambda_{\Phi}}{2\pi} \Phi(\hat{\rho})
 - i (\vec{\bf m}_{2})^{a}_{i} (\vec{\bf b}_{1})^{b}_{j}K_{6}^{ab}
    \epsilon_{kj} H_{ik}(\hat{\rho})
\label{A-pq-2}
\end{align}

 In the case $K_{2} = K_{4} = K_{6} =0$, the angular integration (\ref{A-pq}) 
provides the constrainst 
$
\sum_{a,i} \left( (\vec{\bf m}_{1})^{a}_{i} (\vec{\bf b}_{2})^{a}_{i}
+(\vec{\bf m}_{2})^{a}_{i} (\vec{\bf b}_{1})^{a}_{i}
\right) = 0$ upon fusion, implying that the condition 
$\sum_{a} \vec{\bf m}^{a}. \vec{\bf b}^{a}=0  $ is preserved. 
Unlike the scalar case, this is not sufficient to forbid the generation of 
composite charges. For arbitrary  $K_{2},K_{4},K_{6}$, these composite 
charges will certainly be generated upon coarse-graining.

\subsection{Annihilation of charges : the
screening}\label{sec:screening}

 Now we consider the situation of two opposite charges 
$\vec{\bf b}_{1}+\vec{\bf b}_{2} = \vec{\bf m}_{1}+\vec{\bf m}_{2} =  \vec{0}$.
 The correction to the partition function coming from the configurations with these 
opposite charges still take the form of (\ref{eq:Z-pq}), with the condition  
$\vec{\bf b}_{1}=-\vec{\bf b}_{2};\vec{\bf m}_{1}=-\vec{\bf m}_{2}$. This condition 
implies that the first term of the gradient expansion, considered in 
(\ref{eq:Z-12-fusion}), now only provides a constant term to the free energy, which  
we will neglect. To get the first non-trivial corrections to the system's 
thermodynamics,  we have to consider this gradient expansion up to second order. 
  To this purpose, we expand the action $S_{1,2/\alpha}$ in powers of $\rho$, 
{\it i.e} of $a_{0}$, with charges $\vec{\bf b}_{1/2}, \vec{\bf m}_{1/2}$ 
now  located in $\vec{R}$.   
 In the present case the terms of order $0$ and $2$ vanish as the pair $1,2$ is neutral, 
and we obtain 
\begin{multline} \label{eq:S-12-annihil}
S_{1,2/\alpha}  =  
b_{1,i}^{a}
\rho_{\eta}\partial_{\eta} V_{(1),ij}^{ab}(\vec{R}-\vec{r}_{\alpha}) b_{\alpha,j}^{b} 
+ m_{1,i}^{a}
\rho_{\eta}\partial_{\eta} V_{(3),ij}^{ab}(\vec{R}-\vec{r}_{\alpha}) m_{\alpha,j}^{b} 
\\
 + i \left( m_{1,i}^{a}
\rho_{\eta}\partial_{\eta} \mathcal{G}_{ij}^{ab}(\vec{R}-\vec{r}_{\alpha})
 b_{\alpha,j}^{b} 
-  m_{\alpha,i}^{a} 
\rho_{\eta}\partial_{\eta}\mathcal{G}_{ij}^{ab} (\vec{r}_{\alpha}-\vec{R}) 
b_{1,j}^{b}  \right)
+ \mathcal{O}(a_{0}^{3})
\end{multline}
 Expanding the second exponential to second order in $a_{0}$, 
the correction (\ref{eq:Z-pq}) 
takes the form\footnote{Notice the $\frac12$ factor in front of the sum over 
$(\vec{\bf b}_{1},\vec{\bf m}_{1})$, which accounts for the indiscernability of the 
charges $1$ and $2$.} 
 \begin{multline}
Z_{1,2} = 
\sum_{ \{\vec{\bf b}_{\alpha},\vec{\bf m}_{\alpha}\}, \alpha\neq 1,2}
\left( \prod_{\alpha\neq 1,2} \int \frac{d^{2}\vec{r}_{\alpha}}{a_{0}^{2}} 
 Y[\vec{\bf b}_{\alpha},\vec{\bf m}_{\alpha}]\right)
\frac{1}{2} \sum_{(\vec{\bf b}_{1},\vec{\bf m}_{1})}
 Y[\vec{\bf b}_{1},\vec{\bf m}_{1}]Y[-\vec{\bf b}_{1},-\vec{\bf m}_{1}] 
\\
\int \frac{d^{2}\vec{R}}{a_{0}^{2}}  
\int_{a_{0}\leq |\vec{\rho}| \leq a_{0} e^{dl}} \frac{d^{2}\vec{\rho}}{a_{0}^{2}} 
\left( 
1+\sum_{\alpha}S_{1,2/\alpha}
+\frac{1}{2} \sum_{\alpha,\beta}S_{1,2/\alpha}S_{1,2/\beta}
\right)
e^{\tilde{S}[\vec{b}_{1},\vec{m}_{1}]}
\label{eq:Z-12-annihil}
\end{multline}
 with $S_{1,2/\alpha}$ given by (\ref{eq:S-12-annihil}) and 
$\tilde{S}[\vec{b}_{1},\vec{m}_{1}]$ by (\ref{S-pq}) with 
$\vec{\bf b}_{2}=-\vec{\bf b}_{1}, \vec{\bf m}_{2}=-\vec{\bf m}_{1}$ : 
\begin{multline}\label{def:Stilde}
\tilde{S}[\vec{b}_{1},\vec{m}_{1}] = 
- b_{1,i}^{a}  V_{(1),ij}^{ab}(\vec{\rho}) b_{1,j}^{b} 
- m_{1,i}^{a}  V_{(3)}^{ab}(\vec{\rho}) m_{1,j}^{b} \\
- i ~\left(
  m_{1,i}^{a} \mathcal{G}_{ij}^{ab}( \vec{\rho})  b_{1,j}^{b} 
+ m_{1,i}^{a} \mathcal{G}_{ij}^{ab}(-\vec{\rho})  b_{1,j}^{b} 
\right)
\end{multline}
 As explained above, the first term can be neglected as it renormalizes 
by a constant the free energy. The second term vanishes by the symmetry 
$\vec{\rho} \to -\vec{\rho}$ of the integral. 
 Using $\int d^{2}\vec{\rho}/a_{0}^{2} = dl \int \hat{\rho}$ (where the last integral 
runs over the unit circle), the correction from (\ref{eq:Z-12-annihil}) that we will 
focus on can be written explicitly as\footnote{
Note that similarly to the case of the scalar Coulomb gas, the term 
$\alpha =\beta$ in this sum generates a renormalisation of order $Y^3$ to 
the fugacity 
$Y[\vec{\bf b},\vec{\bf m}]$, which will be neglected in the present study. 
} 
\begin{equation}
Z_{1,2} = \sum_{ \{\vec{\bf b}_{\alpha},\vec{\bf m}_{\alpha}\}, \alpha\neq 1,2}
\left( \prod_{\alpha\neq 1,2} \int \frac{d^{2}\vec{r}_{\alpha}}{a_{0}^{2}} 
 Y[\vec{\bf b}_{\alpha},\vec{\bf m}_{\alpha}]\right)
\frac{1}{2} \sum_{\alpha,\beta}
dS[(\vec{\bf b}_{\alpha},\vec{\bf m}_{\alpha});(\vec{\bf b}_{\beta},\vec{\bf m}_{\beta})]
\end{equation}
 with the (correction to the) action 
\begin{align} \label{eq:int-fusion} 
 dS[&(\vec{\bf b}_{\alpha},\vec{\bf m}_{\alpha}); 
    (\vec{\bf b}_{\beta},\vec{\bf m}_{\beta})]
 = \\ 
\nonumber
&  dl ~ \frac12 \sum_{(\vec{\bf b}_{1},\vec{\bf m}_{1}) } 
 Y[\vec{\bf b}_{1},\vec{\bf m}_{1}]Y[-\vec{\bf b}_{1},-\vec{\bf m}_{1}]
\int d^{2}\vec{R} \int d\hat{\rho}e^{\tilde{S}[\vec{b}_{1},\vec{m}_{1}]}
\hat{\rho}_{s} \hat{\rho}_{t}
\\ \nonumber 
& \biggl[
b_{1,i}^{a} \partial_{s} V_{(1),ij}^{ab} b_{\alpha,j}^{b} 
+ 
m_{1,i}^{a} \partial_{s} V_{(3),ij}^{ab} m_{\alpha,j}^{b} 
+ i \left(
  m_{1,i}^{a} \partial_{s} \mathcal{G}_{ij}^{ab} b_{\alpha,j}^{b} 
+ m_{\alpha,j}^{b} \partial_{s}\mathcal{G}_{ji}^{ba} b_{1,i}^{a}  
\right)
\biggr]
\\ \nonumber
\times &
\biggl[
b_{1,k}^{c} \partial_{t} V_{(1),kl}^{cd} b_{\beta,l}^{d} 
+ 
m_{1,k}^{c} \partial_{t} V_{(3),kl}^{cd} m_{\beta,l}^{d} 
+ i \left(
m_{1,k}^{c} \partial_{t} \mathcal{G}_{kl}^{cd} b_{\beta,l}^{d} 
+ 
m_{\beta,l}^{d} \partial_{t}\mathcal{G}_{lk}^{dc} b_{1,k}^{c}  \right)
\biggl]
\end{align}
 This correction to the action between two charges can be rewritten as 
\begin{align}
\label{contract-monster} 
 dS[ (\vec{\bf b}_{\alpha},\vec{\bf m}_{\alpha}) & 
    ;(\vec{\bf b}_{\beta},\vec{\bf m}_{\beta})]
 = 
\\ \nonumber 
b_{\alpha,j}^{b}b_{\beta,l}^{d}  &
\biggl[
   [M_{1}]_{st,ik}^{ac} [ I^{(1,1)}_{1} ]_{s,ij;t,kl}^{ab;cd} (\vec{r}_{\alpha\beta})
+i [ M_{2} ]_{st,ik}^{ac} [ I^{(1)}_{2} ]_{s,ij;t,kl}^{ab;cd} (\vec{r}_{\alpha\beta})
\\ \nonumber
&
+i [ M_{2}]_{st,ki}^{ca} [ I^{(1)}_{2}]_{t,kl;s,ij}^{cd;ab} (\vec{r}_{\alpha\beta})
-  [ M_{3}]_{st,ik}^{ac}[ I_{3}]_{s,ij;t,kl}^{ab;cd} (\vec{r}_{\alpha\beta})
\biggr]
\\ \nonumber
+ m_{\alpha,j}^{b}m_{\beta,l}^{d} &
\biggl[
   [M_{3}]_{st,ik}^{ac} [ I^{(3,3)}_{1} ]_{s,ij;t,kl}^{ab;cd} (\vec{r}_{\alpha\beta})
+i [ M_{2} ]_{st,ki}^{ca} [ I^{(3)}_{2} ]_{s,ij;t,lk}^{ab,dc} (\vec{r}_{\alpha \beta})
\\ \nonumber
&
+i [ M_{2}]_{st,ik}^{ac}  [ I^{(3)}_{2}]_{t,kl;s,ji}^{cd;ba} (\vec{r}_{\alpha\beta})
- [ M_{1}]_{st,ik}^{ac}[ I_{3}]_{s,ji;t,lk}^{ba;dc} (\vec{r}_{\alpha\beta})
\biggr]
\\ \nonumber 
+ b_{\alpha,j}^{b}m_{\beta,l}^{d}
& \biggl[
   [ M_{2} ]_{st,ik}^{ac}[ I^{(1,3)}_{1} ]_{s,ij;t,kl}^{ab;cd} (\vec{r}_{\alpha\beta})
+i [ M_{1} ]_{st,ik}^{ac}[ I^{(1)}_{2} ]_{s,ij;t,lk}^{ab,dc} (\vec{r}_{\alpha \beta})
\\ \nonumber
&
+i [ M_{3} ]_{st,ik}^{ac}[ I^{(3)}_{2} ]_{t,kl;s,ij}^{cd,ab} (\vec{r}_{\alpha\beta})
-  [ M_{2} ]_{st,ki}^{ca}[ I_{3}]_{s,ij;t,lk}^{ab;dc} (\vec{r}_{\alpha\beta})
\biggr]
\\ \nonumber 
+ m_{\alpha,j}^{b}b_{\beta,l}^{d}
& \biggl[
   [ M_{2} ]_{st,ki}^{ca}[ I^{(3,1)}_{1} ]_{s,ij;t,kl}^{ab;cd} (\vec{r}_{\alpha\beta})
+i [ M_{1} ]_{st,ik}^{ac}[ I^{(1)}_{2} ]_{t,kl;s,ji}^{cd,ba} (\vec{r}_{\alpha \beta})
\\ \nonumber
&
+i [ M_{3} ]_{st,ik}^{ac}[ I^{(3)}_{2} ]_{s,ij;t,kl}^{ab,cd} (\vec{r}_{\alpha\beta})
-  [ M_{2} ]_{st,ik}^{ac}[ I_{3}]_{s,ji;t,kl}^{ba;cd} (\vec{r}_{\alpha\beta})
\biggr]
\end{align}
 where we define the tensors relative respectively to the integration over 
 $\vec{\rho}$ and $\vec{R}$ : 

\begin{subequations}
\label{def-Ms}
\begin{align}\label{def-M1} 
& [M_{1}]_{st,ik}^{ac}  =  
\frac{dl}{2} 
\sum_{(\vec{\bf b}_{1},\vec{\bf m}_{1}) }  
 Y[\vec{\bf b}_{1},\vec{\bf m}_{1}]Y[-\vec{\bf b}_{1},-\vec{\bf m}_{1}]
\int d \hat{\rho}
~ e^{\tilde{S}[\vec{b}_{1},\vec{m}_{1}]}
\hat{\rho}_{s}\hat{\rho}_{t}
b_{1,i}^{a} b_{1,k}^{c} 
         \\  \label{def-M2}
& [ M_{2} ]_{st,ik}^{ac}  =
\frac{dl}{2} 
\sum_{(\vec{\bf b}_{1},\vec{\bf m}_{1}) }  
 Y[\vec{\bf b}_{1},\vec{\bf m}_{1}]Y[-\vec{\bf b}_{1},-\vec{\bf m}_{1}]
\int d \hat{\rho}
~ e^{\tilde{S}[\vec{b}_{1},\vec{m}_{1}]}
\hat{\rho}_{s}\hat{\rho}_{t}
 b_{1,i}^{a} m_{1,k}^{c}
\\
& [ M_{3} ]_{st,ik}^{ac}  =
\frac{dl}{2} 
\sum_{(\vec{\bf b}_{1},\vec{\bf m}_{1}) }  
 Y[\vec{\bf b}_{1},\vec{\bf m}_{1}]Y[-\vec{\bf b}_{1},-\vec{\bf m}_{1}]
\int d \hat{\rho}
~ e^{\tilde{S}[\vec{b}_{1},\vec{m}_{1}]}
\hat{\rho}_{s}\hat{\rho}_{t}
m_{1,i}^{a} m_{1,k}^{c}
\end{align}
\end{subequations}
\begin{subequations}
\label{def-Is}
\begin{align}
 \label{def-I1}
& [ I^{(1,3)}_{1} ]_{s,ij;t,kl}^{ab;cd} (\vec{r}_{\alpha}-\vec{r}_{\beta})  = 
\int d^{2}\vec{R} ~
\partial_{s} V^{ab}_{(1),ij} (\vec{R}-\vec{r}_{\alpha})
\partial_{t} V^{cd}_{(3),kl} (\vec{R}-\vec{r}_{\beta})
\\ \label{def-I2}
& [ I^{(1)}_{2} ]_{s,ij;t,kl}^{ab;cd} (\vec{r}_{\alpha}-\vec{r}_{\beta})  = 
\int d^{2}\vec{R} ~
\partial_{s} V^{ab}_{(1),ij} (\vec{R}-\vec{r}_{\alpha})
\partial_{t}  \mathcal{G}^{cd}_{kl} (\vec{R}-\vec{r}_{\beta})
\\
& [ I_{3}]_{s,ij;t,kl}^{ab;cd} (\vec{r}_{\alpha}-\vec{r}_{\beta})  =  
\int d^{2}\vec{R} ~
\partial_{s} \mathcal{G}^{ab}_{ij} (\vec{R}-\vec{r}_{\alpha})
\partial_{t} \mathcal{G}^{cd}_{kl} (\vec{R}-\vec{r}_{\beta})
\end{align}
\end{subequations}

 where all the above expressions are symetric in $\alpha,\beta$. We have used that all derivatives
of the potentials $V$ and $\mathcal{G}$ are odd.

 To proceed, we thus have to 
 (i) perform the 
  integral over $\vec{R}$,, {\it i.e} calculate explicitly the tensors $I_{1,2,3}$
 (ii) perform the integral over $\vec{\rho}$, 
{\it i.e} calculate explicitly the tensors $M_{1,2,3}$,
 and finally (iii) contract all the tensors in (\ref{contract-monster}). 
 If this final contraction can be cast into contributions to the 
initial potential $V_{ij}^{ab}(\vec{r}),\mathcal{G}_{ij}^{ab}(\vec{r})$, this 
will prove the renormalizability of the present vector Coulomb gaz to one loop. 

\subsubsection{Integration over $\vec{R}$}

 We focus on the tensors $I_{1,2,3}$, which are all integral of double products 
of gradients of $V,\mathcal{G}$. 
These integrations are conveniently done in Fourier space, and we start by obtaining 
Fourier representation of these potential's gradients : with the 
definition of the projectors 
$P_{ij}^L (\hat{q}) = \hat{q}_i \hat{q}_j$ and 
$P_{ij}^T (\hat{q}) = \delta_{ij} - \hat{q}_i \hat{q}_j 
= \epsilon_{ik} \epsilon_{jl}  P_{kl}^L (\hat{q})$, we obtain, from the 
definition 
\begin{equation}
V_{ij}^{ab}(K_1,K_2)(\vec{r}) = 
\int \frac{d^2 \vec{q}}{(2\pi)^2} 
\left(1-e^{i~\vec{q}.\vec{r}} \right) \frac{2 \pi }{q^2} 
\left[ (K_1-K_2)^{ab} ~P_{ij}^L + ( K_1+K_2)^{ab}~P_{ij}^T \right] 
\end{equation} 
 the expression or its gradient 
\begin{align} \nonumber
\partial_{s} V_{(1),ij}^{ab}(\vec{r}) & =  
-2\pi i \left[ (K_1-K_2)^{ab} \delta_{ik} \delta_{jl} 
+ (K_1+K_2)^{ab} \epsilon_{ik}\epsilon_{jl} \right]
\int \frac{d^2 \vec{q} }{(2 \pi)^2} ~ e^{i \vec{q}.  \vec{r} } \frac{q_{s}}{q^2} 
	 P_{kl}^L(\hat{q}) \\
& \equiv  - 2\pi i~ \mathcal{C}_{ijkl}^{ab}(K_1,K_2) 
\int \frac{d^2 \vec{q} }{(2 \pi)^2}~ e^{i \vec{q}.\vec{r} } 
	 \frac{q_{s}q_{k}q_{l}}{q^4}
\end{align}
 Similarly, using the equality (\ref{eq:G-V}) the second gradient 
reads 
\begin{align}
\partial_{s} \mathcal{G}^{ab}_{ij}(\vec{r}) & =
\frac{-\lambda_{\Phi}}{2\pi}
\delta^{ab}\epsilon_{s t} \partial_{t} V_{ij}(1,0) + 
\epsilon_{mj} \partial_{s} V_{im}^{ab}(K_5,K_6)   \\
& =  - 2i \pi \left[ 
\frac{-\lambda_{\Phi}}{2\pi}
\delta^{ab}\epsilon_{s t} \mathcal{C}_{ijkl} (1,0) + 
\epsilon_{mj} \delta_{s t} \mathcal{C}_{imkl}^{ab}(K_5,K_6) 
	 \right] 
\int \frac{d^2 \vec{q} }{(2 \pi)^2} ~ e^{i \vec{q}.\vec{r}} 
\frac{q_{t}}{q^2} P_{kl}^L \\ \label{eq:intermediate}
& \equiv  -2 i D_{ ijkl,st }^{ab} 
\int \frac{d^2 \vec{q}}{(2 \pi)^2} ~ e^{i \vec{q}.\vec{r} } \frac{q_{t}q_{k}q_{l}}{q^{4}}
\end{align}
 where we have defined
\begin{align}
\label{eq:C}
\mathcal{C}_{ijkl}^{ab}(K_{1},K_{2}) & =  
(K_1-K_2)^{ab} \delta_{ik} \delta_{jl} + 
(K_1 + K_2)^{ab} ( \delta_{ij}\delta_{kl}-\delta_{il}\delta_{jk}) 
\\
\label{eq:D}
D_{ijkl,st}^{ab} & =  
-\frac{\lambda_{\Phi}}{2\pi} \delta^{ab}\epsilon_{st} [ \delta_{ik} 
\delta_{jl} + \delta_{ij}\delta_{kl}-\delta_{il}\delta_{jk} ] \\ 
\nonumber & +\epsilon_{mj} \delta_{st} \left[ 
  (K_5-K_6)^{ab} \delta_{ik} \delta_{ml} 
+ (K_5 + K_6)^{ab} (\delta_{im}\delta_{kl}-\delta_{il}\delta_{mk}) \right]
\end{align}
 With this representation, the integrals $I_{1,2,3}$ are expressed as 
\begin{align}
\label{eq:I1}
[ I^{(1,3)}_{1} ]_{s,ij;t,kl}^{ab;cd} (\vec{r}_{\alpha}-\vec{r}_{\beta})  & =
-4\pi^2 \mathcal{C}_{ijmn}^{ab}(K_1,K_2)   \mathcal{C}_{ klpq}^{cd} (K_3,K_4)
Q_{mnpqst}(\vec{r}_{\alpha}-\vec{r}_{\beta}) \\
\label{eq:I2}
[ I^{(1)}_{2} ]_{s,ij;t,kl}^{ab;cd} (\vec{r}_{\alpha}-\vec{r}_{\beta}) & =
-4 \pi^2 \mathcal{C}_{ijmn}^{ab}(K_1,K_2)   D_{klpq,tu}^{cd} 
Q_{mnpqsu}(\vec{r}_{\alpha}-\vec{r}_{\beta}) \\
\label{eq:I3}
[ I_{3}]_{s,ij;t,kl}^{ab;cd} (\vec{r}_{\alpha}-\vec{r}_{\beta})  & = 
- 4 \pi^2  D_{ijmn,su}^{ab}   D_{klpq,tv}^{cd} 
Q_{mnpq,uv}(\vec{r}_{\alpha}-\vec{r}_{\beta}) 
\end{align}
 where we have defined the integral 
\begin{align} \nonumber
Q_{klmnst} (\vec{r}_{\alpha}-\vec{r}_{\beta})  & = 
\int d^2\vec{R} \int \frac{d^2 \vec{q}}{(2 \pi)^2} \int \frac{d^2 \vec{q'}}{(2 \pi)^2}
e^{i \left( \vec{q}.(\vec{R}-\vec{r}_{\alpha})+ \vec{q'}.(\vec{R}-\vec{r}_{\beta}) \right) }
\frac{q_{s} q_k q_l q'_{t} q'_m q'_n}{q^{4} (q')^{4}} \\
& =  
\frac{ \partial^6 }{
\partial_{s} \partial_{t} \partial_{k} \partial_{l} \partial_{m} \partial_{n}} \int 
\frac{d^2 \vec{q}}{(2 \pi)^2} 
\frac{1}{q^8} e^{i \vec{q}.(\vec{r}_{\alpha}-\vec{r}_{\beta})}
\end{align}
 Using the Schwinger representation, the last integral yields :
\begin{equation}
\int \frac{d^2 \vec{q}}{(2 \pi)^2} \frac{1}{q^8} e^{i \vec{q}.\vec{r}} 
 =  \frac{1}{6} \int \frac{d^2 \vec{q}}{(2 \pi)^2} 
 \int_0^{\infty} d u u^3 e^{-u q^2 + i \vec{q}.\vec{r}} 
=
\frac{1}{12} \int_0^{\infty} \frac{du}{2 \pi u} u^3 e^{- \frac{r^2}{4 u}}
\end{equation}
 The differenciation of the gaussian up to order 6 is now straigthforward :
\begin{align} \nonumber
\frac{\partial^6 }
{\partial_{s} \partial_{t} \partial_{k} \partial_{l} \partial_{m} \partial_{n}} 
\left( e^{-\frac{A}{2} r^2} \right) 
= &  
\biggl[
-A^3 \delta_{st} \delta_{kl} \delta_{mn} 
+ c.p.  {\rm (15 terms)} 
\\ \nonumber
& + A^4 ~ r_{s}r_{t} \delta_{kl} \delta_{mn} + c.p.  {\rm (45 terms)}
\\ \nonumber  
& - A^5 ~ r_{s}r_{t} r_k r_l \delta_{mn} + c.p.  {\rm (15 terms)} 
\\ \label{eq:schwinger}
& + A^6 ~ r_{s}r_{t} r_k r_l r_m r_n
\biggr]  e^{-\frac{A}{2} r^2}
\end{align}
 where $c.p.$ means circular permutation of the indices (the 
number of corresponding permutated terms is indicated). 
Finally, using $ \int_0^{\infty} du u^{\beta} e^{-u} = \Gamma(\beta+1) $, 
 we find  :
\begin{align} \nonumber
12 \times Q_{stklmn} (\vec{r})   = 
& \frac{1}{16 \pi} Ei\left( - \frac{r^2}{4 L^2} \right) 
\delta_{st} \delta_{kl} \delta_{mn} 
 + \frac{1}{8 \pi} ~ \hat{r}_{s}\hat{r}_{t} \delta_{kl} \delta_{mn} .  
\\ \label{eq:Q}
& - \frac{1}{4 \pi} ~ \hat{r}_{s}\hat{r}_{t} \hat{r}_k \hat{r}_l \delta_{mn}  
 + \frac{1}{\pi} ~ \hat{r}_{s}\hat{r}_{t} \hat{r}_k \hat{r}_l \hat{r}_m \hat{r}_n
+(c.p.)
\end{align}

In this expression $L$ stands for an IR cut-off.  We will use the following asymptotic limit for the exponential 
integral\cite{carpentier97} : $ Ei(-x) \simeq 
\gamma + \ln(x)$ in the limit $x\rightarrow 0$.  
 The expressions (\ref{eq:C},\ref{eq:D},\ref{eq:Q}), together with the contractions 
formula (\ref{eq:I1},\ref{eq:I2},\ref{eq:I3}) constitute our final explicit 
expressions for the integrals $I_{1,2,3}$.

\subsubsection{Integration over $\vec{\rho}$}

 The invariance under $2\pi /3$ rotations of the integrals in $M_{1,2,3}$, 
defined in eq. (\ref{def-Ms}), ensures that these 
tensors are isotropic, provided that the fugacity of a vector charge 
$Y[\vec{\bf b},\vec{\bf m}]$ is constant under any rotation of the charge 
$\vec{\bf b},\vec{\bf m}$ (in particular, this implies 
$Y[\vec{\bf b},\vec{\bf m}]= Y[-\vec{\bf b},-\vec{\bf m}]$). 
 Using this isotropy, we decompose the tensors $M_{1,2,3}$ according to\footnote{
 The tensor $U$ and $\tilde{U}$ arises as can be seen e.g. by expanding the definition of $M^{(2)}$ to first order 
 in $K_{6}$ which yields tensors of the form (in the case $m.b=0$)
 $$
\sum_{b,m}\int d\rho  \hat{\rho}_{s}\hat{\rho}_{t}b_{i}m_{k}
\left( 
(\rho.m) (\rho.b^\perp) -\frac12 m.b^\perp
\right)
 $$ }  
\begin{align}\label{def:Gamma}
[M_{w}]_{st,ik}^{ac}  & = 
dl \left(
\left(\Gamma_w^{ac}-\tilde{\Gamma}_w^{ac}\right) T_{st,ik} +  \tilde{\Gamma}_w^{ac} \tilde{T}_{st,ik} \right)
\quad ; \quad w=1,3 
\\
i [M_{2}]_{st,ik}^{ac}  & = 
 dl \left(
\left(\Gamma_2^{ac}-\tilde{\Gamma}_2^{ac}\right)  U_{st,ik} +  \tilde{\Gamma}_2^{ac} \tilde{U}_{st,ik} \right)
\end{align}
 and where we used the definitions of the symetric and antisymetric tensors  
\begin{align}
& T_{st,ik} = \delta_{st} \delta_{ik} 
\quad ; \quad 
 \tilde{T}_{st,ik} = \delta_{si} \delta_{tk}+  \delta_{ti} \delta_{sk}
\\
& U_{st,ik} = \delta_{st} \epsilon_{ik} 
\quad ; \quad 
\tilde{U}_{st,ik} = \delta_{sk} \epsilon_{it}+ \delta_{tk} \epsilon_{is}
\end{align}
By using (note the unusual definition of the trace) : 
\begin{align}
& \textrm{Tr} (AB)\equiv \sum_{st,ik}A_{st,ik}B_{st,ik} ,\\
& \textrm{Tr} (T^2)=4,  \textrm{Tr} ( T \tilde{T}) = 4, 
  \textrm{Tr} (\tilde{T}\tilde{T}) = 12, \\
& \textrm{Tr} ( U^2) = 4 , 
\textrm{Tr} ( \tilde{U}^2) = 12 ,  
\textrm{Tr} ( U \tilde{U}) = 4  , 
\end{align}
 we obtain the formal expression for the coefficients $\Gamma_w^{ac},\tilde{\Gamma}_w^{ac}$ : 
\begin{subequations} \label{eq:Gamma} 
\begin{align} 
& dl \Gamma_{w}^{ac} = 
\frac{1}{4} \textrm{Tr} (TM_w^{ac}) 
\quad ; \quad w=1,3 
\\
& dl \tilde{\Gamma}_{w}^{ac}= 
-\frac{1}{8} \textrm{Tr} (TM_w^{ac})  +\frac{1}{8}  \textrm{Tr}(\tilde{T}M_w^{ac}) 
\quad ; \quad w=1,3 
\\
& dl \Gamma_{2}^{ac} = 
\frac{i}{4} \textrm{Tr} (UM_2^{ac}) 
\\
& dl \tilde{\Gamma}_{2}^{ac}= 
-\frac{i}{8} \textrm{Tr} (UM_2^{ac})  +\frac{i}{8}  \textrm{Tr}(\tilde{U}M_2^{ac}) 
\end{align}
\end{subequations}

 with 
\begin{subequations} \label{eq:TrTM}
\begin{align}
\textrm{Tr} (TM_{1}^{ac}) &= 
\frac{dl}{2}
\sum_{(\vec{\bf b},\vec{\bf m}) }  
 Y^{2}[\vec{\bf b},\vec{\bf m}]
(\vec{b}^{a}.\vec{b}^{c})
\int d \hat{\rho}
~ e^{\tilde{S}[\vec{b},\vec{m}]}
\\
\textrm{Tr}(\tilde{T}M_{1}^{ac}) &= 
dl
\sum_{(\vec{\bf b},\vec{\bf m}) }  
 Y^{2}[\vec{\bf b},\vec{\bf m}]
\int d \hat{\rho}
~ e^{\tilde{S}[\vec{b},\vec{m}]}
 (\hat{\rho}.\vec{b}^{a})
(\hat{\rho}.\vec{b}^{c})
\\
\textrm{Tr} (UM_{2}^{ac}) &= 
-\frac{dl}{2}
\sum_{(\vec{\bf b},\vec{\bf m}) }  
 Y^{2}[\vec{\bf b},\vec{\bf m}]
(\vec{b}^{a,\perp}.\vec{m}^{c})
\int d \hat{\rho}
~ e^{\tilde{S}[\vec{b},\vec{m}]}
\\
\textrm{Tr}(\tilde{U}M_{2}^{ac}) &= 
-dl
\sum_{(\vec{\bf b},\vec{\bf m}) }  
 Y^{2}[\vec{\bf b},\vec{\bf m}]
\int d \hat{\rho}
~ e^{\tilde{S}[\vec{b},\vec{m}]}
\left[ 
(\hat{\rho}.\vec{b}^{a,\perp}) (\hat{\rho}.\vec{m}^{c})
\right]
\\
\textrm{Tr} (TM_{3}^{ac}) &= 
\frac{dl}{2}
\sum_{(\vec{\bf b},\vec{\bf m}) }  
 Y^{2}[\vec{\bf b},\vec{\bf m}]
(\vec{m}^{a}.\vec{m}^{c})
\int d \hat{\rho}
~ e^{\tilde{S}[\vec{b},\vec{m}]}
\\
\textrm{Tr}(\tilde{T}M_{3}^{ac}) &= 
dl
\sum_{(\vec{\bf b},\vec{\bf m}) }  
 Y^{2}[\vec{\bf b},\vec{\bf m}]
\int d \hat{\rho}
~ e^{\tilde{S}[\vec{b},\vec{m}]}
 (\hat{\rho}.\vec{m}^{a})
(\hat{\rho}.\vec{m}^{c})
\end{align}
\end{subequations}

 Note the following useful relations :
\begin{align}
\textrm{Tr}(\tilde{T}M_{1}^{ac})  - \textrm{Tr}(TM_{1}^{ac}) & = 
\frac{\partial}{\partial K_{2}^{ac} }
\left( 
dl
\sum_{(\vec{\bf b},\vec{\bf m}) }  
 Y^{2}[\vec{\bf b},\vec{\bf m}]
\int d \hat{\rho}
~ e^{\tilde{S}[\vec{b},\vec{m}]}
\right) 
\\
\textrm{Tr}(\tilde{T}M_{3}^{ac})  - \textrm{Tr}(TM_{3}^{ac}) & = 
\frac{\partial}{\partial K_{4}^{ac} }
\left( 
dl
\sum_{(\vec{\bf b},\vec{\bf m}) }  
 Y^{2}[\vec{\bf b},\vec{\bf m}]
\int d \hat{\rho}
~ e^{\tilde{S}[\vec{b},\vec{m}]}
\right) 
\\
\textrm{Tr}(\tilde{U}M_{2}^{ac})  - \textrm{Tr}(UM_{2}^{ac}) & = 
- \frac{\partial}{\partial K_{6}^{ac} }
\left( 
\frac{dl}{2}
\sum_{(\vec{\bf b},\vec{\bf m}) }  
 Y^{2}[\vec{\bf b},\vec{\bf m}]
\int d \hat{\rho}
~ e^{\tilde{S}[\vec{b},\vec{m}]}
\right) 
 \end{align}

\subsubsection{Final contraction of tensors}

With the above expressions for the $M$ and $I$ tensors, we can now explicitly perform 
the contractions of eq. (\ref{contract-monster}). This tedious task is performed using mathematica.
We find that the result can be cast in the same form as the original interaction
with changes $dK_i$ in the couplings : this proves the renormalizability of the model to order 
$Y^2$. Additional constants are produced which correct fugacities to cubic order in $Y$. 
The result of these contractions is presented in the following section.

\section{Resulting RG equations for the general model} 
\label{sec:RG-general}

In this Section we collect and analyze the RG equations for the
fugacity variables $Y[\vec {\bf b}, \vec {\bf m}]$ and the matrices $K_i$, $i=1,..6$, which parameterize the
general VECG model defined by the action (\ref{eq:S-VECG}).

\subsection{Scaling equations for the fugacities}

The equations (\ref{eq:repara},\ref{RG-fusion}) provide the full equations for the fugacities : 
\begin{multline}
\partial_l Y[\vec{\bf b},\vec{\bf m}] = 
\left( 2-
\frac12
\left( 
\vec{b}^{a}.\vec{b}^{b}  K_{1}^{ab} + 
\vec{m}^{a}.\vec{m}^{b}  K_{3}^{ab}
+2 i m^{a}_{i} \epsilon_{ij} b_{j}^{b} K_{5}^{ab}
\right)\right)Y[\vec{\bf b},\vec{\bf m}]
\\
+  
\sum_{(\vec{\bf b}_{1},\vec{\bf m}_{1})+(\vec{\bf b}_{2},\vec{\bf m}_{2})= (\vec{\bf b},\vec{\bf m})}
A_{(\vec{\bf b}_{1},\vec{\bf m}_{1});(\vec{\bf b}_{2},\vec{\bf m}_{2})} 
Y[\vec{\bf b}_{1},\vec{\bf m}_{1}]Y[\vec{\bf b}_{2},\vec{\bf m}_{2}]
\end{multline}
where the numerical factor 
$A_{(\vec{\bf b}_{1},\vec{\bf m}_{1});(\vec{\bf b}_{2},\vec{\bf m}_{2})} $
is defined in eqs. (\ref{A-pq}) and (\ref{A-pq-2}) 
\begin{equation}
A_{(\vec{\bf b}_{1},\vec{\bf m}_{1});(\vec{\bf b}_{2},\vec{\bf m}_{2})}= 
\int d\hat{\rho} 
\exp(S[(\vec{\bf b}_{1},\vec{\bf m}_{1});(\vec{\bf b}_{2},\vec{\bf m}_{2})])
\end{equation}
 with the action $S[(\vec{\bf b}_{1},\vec{\bf m}_{1});(\vec{\bf b}_{2},\vec{\bf m}_{2})]$ given 
by (\ref{S-pq}) with $\rho = a_{0}$ :  
\begin{align} \nonumber
S[(\vec{\bf b}_{1},\vec{\bf m}_{1});(\vec{\bf b}_{2},\vec{\bf m}_{2})] = &
- K_{2}^{ab}  \vec{\bf b}_{1}^{a}.H(\hat{\rho}).   \vec{\bf b}_{2}^{b} 
- K_{4}^{ab}  \vec{\bf m}_{1}^{a}.H(\hat{\rho}).  \vec{\bf m}_{2}^{b}
\\ \nonumber
&+ i (  \vec{\bf m}_{1}^{a}. \vec{\bf b}_{2}^{a} + \vec{\bf m}_{2}^{a}. \vec{\bf b}_{1}^{a})
       \frac{\lambda_{\Phi}}{2\pi} \Phi(\hat{\rho})
\\
&-i K_{6}^{ab} \left( \vec{\bf m}_{1}^{a}.H(\hat{\rho}).  \vec{\bf b}_{2}^{b\perp}
             +\vec{\bf m}_{2}^{a}.H(\hat{\rho}).  \vec{\bf b}_{1}^{b\perp} \right)
\end{align}

The evaluation of the coefficients $A_{(\vec{\bf b}_{1},\vec{\bf m}_{1});(\vec{\bf b}_{2},\vec{\bf m}_{2})} $
is model dependent. For the models considered here, it will be performed in 
subsequent publication. 
 
\subsection{Scaling equations for the couplings matrices}
  
 The RG equations for the coupling constants $K_{i}$ are obtained by performing the tensors contractions of 
 eq.~(\ref{contract-monster}). The resulting expression is displayed in the appendix \ref{sec:fullRG}.
Here we show that their structure can be further simplified by introducing the new couplings $p_{i}$ defined by:
\begin{align}
& p_1  = 2\pi \left(  K_1+K_2 \right)
\quad ; \quad 
p_2  = 2\pi \left( K_1-K_2  \right)
\\
& p_3  =  2\pi \left(K_3 +K_4  \right)
\quad ; \quad 
p_4  =  2\pi \left(K_3-K_4  \right) \\
& p_5  =  2\pi \left(K_5+K_6 \right) 
\quad ; \quad 
p_6  =  2\pi \left(K_5-K_6  \right) . 
\end{align}
In the general case the $p_i$ (and the $\Gamma_i$ and $\tilde \Gamma_i$ below)
are commuting replica matrices. 
Quite remarkably, the $6$ scaling equations (\ref{eq:RGfull}) decouple into two independent set of 
 $3$ equations for the groups $p_{1},p_{4},p_{6}$, and $p_{2},p_{3},p_{5}$:
\begin{subequations}
\label{RGeq-p-1}
\begin{align}
\partial_{l}p_{1}    = & -\Gamma_{1} p_{1}^2  + \tilde{\Gamma}_{1} p_{1}^2
				+ 2 \Gamma_{2} p_{1}p_{6}  +2  \tilde{\Gamma}_{2} p_{1}(\lambda_{\phi}+p_{6})
				\nonumber \\
				&+ \Gamma_{3} (\lambda_{\phi}^2+p_{6}^2) 
				+  \tilde{\Gamma}_{3} ( \lambda_{\phi}+p_{6})^2
 \\
\partial_{l}p_{4}    = &   +\Gamma_{1} (\lambda_{\phi}^2+p_{6}^2)
				 - \tilde{\Gamma}_{1} (\lambda_{\phi}- p_{6})^2 
				+ 2 \Gamma_{2} p_{4}p_{6} - 2 \tilde{\Gamma}_{2} (\lambda_{\phi}-p_{6})p_{4}
				\nonumber \\
				&- \Gamma_{3}p_{4}^2  -  \tilde{\Gamma}_{3} p_{4}^2
 \\
\partial_{l}p_{6}    = & - \Gamma_{1} p_{1}p_{6}  +  \tilde{\Gamma}_{1} p_{1}(p_{6}-\lambda_{\phi}) 
				+ (\Gamma_{2}+\tilde{\Gamma}_{2}) (-\lambda_{\phi}^2+p_{6}^2-p_{1}p_{4})
				\nonumber\\
&				- \Gamma_{3}p_{4}p_{6}  -  \tilde{\Gamma}_{3} p_{4}(\lambda_{\phi}+p_{6})		
\end{align}
\end{subequations}
 and		
\begin{subequations}
\label{RGeq-p-2}
\begin{align}
\partial_{l}p_{2}    = &  - \Gamma_{1}p_{2}^2  -   \tilde{\Gamma}_{1} p_{2}^2
				+ 2 \Gamma_{2} p_{5}p_{2}  -  2 \tilde{\Gamma}_{2}(\lambda_{\phi}+p_{5})p_{2}
				\nonumber \\
				&+ \Gamma_{3}(\lambda_{\phi}^2+p_{5}^2)  
				- \tilde{\Gamma}_{3} (\lambda_{\phi}+p_{5})^2
 \\
\partial_{l}p_{3}    = &  +\Gamma_{1} (\lambda_{\phi}^2+p_{5}^2)  
				+  \tilde{\Gamma}_{1} (\lambda_{\phi}^2-p_{5})^2
				+ 2 \Gamma_{2}p_{3}p_{5}  + 2 \tilde{\Gamma}_{2}(\lambda_{\phi}-p_{5})p_{3}
				\nonumber \\
				&- \Gamma_{3}p_{3}^2 +  \tilde{\Gamma}_{3}p_{3}^2
 \\
\partial_{l}p_{5}    = &  -\Gamma_{1}p_{2}p_{5}  +  \tilde{\Gamma}_{1} p_{2}(\lambda_{\phi}-p_{5})
				+ (\Gamma_{2}-  \tilde{\Gamma}_{2} )(-\lambda_{\phi}^2+p_{5}^2-p_{2}p_{3})  
\nonumber\\
&				- \Gamma_{3}p_{3}p_{5}  +  \tilde{\Gamma}_{3}(\lambda_{\phi}+p_{5})p_{3}. 
\end{align}
\end{subequations}
where the $\Gamma_i$ and $\tilde \Gamma_i$ were defined in (\ref{eq:Gamma}, \ref{eq:TrTM}).  
Their flow equation can be deduced from the fugacity RG equation given in the
previous section.

In addition these equations possess remarkable symmetries. The following transformation:
\begin{subequations}\label{sym:rotation}
\begin{align}
& p_{1}     \leftrightarrow p_{2}, p_{3} \leftrightarrow p_{4}, p_{5}     \leftrightarrow p_{6} \\  
& \Gamma_{i} \to  \Gamma_{i}  ;\tilde{\Gamma}_{i}    \to  -\tilde{\Gamma}_{i}, i=1,\dots 3. 
\end{align}
\end{subequations}
exchanges these two groups. In terms of the 
Coulomb gas couplings, it corresponds to 
$K_{2}\to -K_{2} ; K_{4}\to -K_{4} ; K_{6}\to -K_{6}$. It can be viewed formally as 
a $\pi/2$ charge rotation $(\vec{b},\vec{m}) \to (\vec{b}^{\perp},\vec{m}^{\perp})$
in the original action. This means that a model where the signs of $K_2,K_4,K_6$
are simultaneously changed is the same (up to an immaterial global rotation) with
the same fugacities.

The second symmetry is the previously discussed electromagnetic
duality. It operates inside each of these groups, i.e the 
RG equations are invariant under:
\begin{subequations}\label{sym:duality}
\begin{align}
& p'_{1} = p_{4}\quad ; \quad
 p'_{4} = p_{1} \quad ; \quad
 p'_{6} = - p_{6} 
\\
& p'_{2}=p_{3} \quad ; \quad
p'_{3}=p_{2} \quad ; \quad
p'_{5}= -p_{5} 
\\
& \Gamma_{1}' = \Gamma_{3} ; \ 
\tilde{\Gamma}_{1}' = - \tilde{\Gamma}_{3} ; \ 
\Gamma_{2}' = - \Gamma_{2} ; \ 
\tilde{\Gamma}_{2}' = - \tilde{\Gamma}_{2} ; \ 
\Gamma_{3}' = \Gamma_{1} ; \ 
\tilde{\Gamma}_{3}' = - \tilde{\Gamma}_{1}
\end{align}
\end{subequations}

\section{Resulting RG equations for the Elastic Models} 
\label{sec:RG-elastic}

We now focus on the models defined at the beginning of the paper,
i.e. an elastic lattice with dislocations in presence of
a substrate, which can include a periodic modulation and/or
a substrate with quenched disorder. At the bare level these models
do not span the whole space of the six $K_i$ (considered in the
previous Section) but only a ''3 dimensional'' subspace of Coulomb gases
(called below the  ''elastic sub-manifold''). 
Indeed these models correspond to the same definitions 
 (\ref{replicaCG}) of the couplings constants $K_{i}$ (resp. replica matrices) in terms of the 
 elastic constants (resp. matrices) $c_{11},c_{66},\gamma$. 
We find, and this is one of the main results of the paper, that
this sub-manifold, i.e. the definitions (\ref{replicaCG}), is preserved
by the RG flow. We emphasize that this property is far from obvious, and cannot be
easily inferred from the structure of the RG equations (\ref{eq:RGfull}) of the full Coulomb gas, without any knowledge of the definitions (\ref{replicaCG}). 

\subsection{Stable elastic sub-manifold}

Let us start by expressing the coupling constants/matrices $p_{i}$ in terms of the elastic constants/matrices. 
The constants from the first group read 
\begin{align}
& p_{1} = 2\pi (K_{1}+K_{2}) = 
\frac{4 a_{0}^{2}}{T} c_{66} (c_{11}-c_{66}) c_{11}^{-1} , 
\\
& p_{4} = 2\pi (K_{3} - K_{4}) = T|\vec{G}_{1}|^{2} c_{11}^{-1} , 
\\
& p_{6} = 2\pi (K_{5} - K_{6}) = 
 a_{0}|\vec{G}_{1}|  \left( 2 \frac{c_{66}}{c_{11}} -1 \right) . 
\end{align}
 Note that these $3$ constants depend only on $c_{11},c_{66}$, and not on $\gamma$. 
 These equations can be inverted into 
\begin{align}
& c_{11} = T |\vec{G}_{1}|^2  p_{4}^{-1}
\quad ; \quad 
 c_{66} = \frac{T |\vec{G}_{1}|}{2 a_{0}} \frac{p_{1}}{\lambda_{\phi}-p_{6}}
\label{invert1}. 
\end{align}
We recall that $\lambda_{\phi} = a_{0} |\vec{G}_{1}|$.

The scaling of $\gamma$ (together with $c_{66}$) is described by the second group of couplings : 
\begin{align}
& p_{2} = 2\pi (K_{1}-K_{2}) = 
\frac{4 a_{0}^{2}}{T} c_{66}  \gamma (c_{66} + \gamma)^{-1} , 
\\
& p_{3} = 2\pi (K_{3} + K_{4}) = T|\vec{G}_{1}|^{2} (c_{66} +\gamma)^{-1} , 
\\
& p_{5} = 2\pi (K_{5} + K_{6}) = 
 a_{0}|\vec{G}_{1}| (c_{66}-\gamma )( c_{66}+\gamma )^{-1}, 
\end{align}
which are inverted into 
 \begin{equation}
 c_{66} = \frac{T |\vec{G}_{1}|}{2 a_{0}} \frac{\lambda_{\phi}+p_{5}}{p_{3}}, \label{invert2} \\
\quad ; \quad 
 \gamma = \frac{T |\vec{G}_{1}|}{2 a_{0}} \frac{\lambda_{\phi}-p_{5}}{p_{3}}.
\end{equation}

From these considerations we find the equations defining the ''elastic sub-manifold''.
\begin{subequations} \label{manifold} 
\begin{align}
\lambda_{\phi}^2 - p_{6}^2 =  p_{1} p_{4}. \\
\lambda_{\phi}^2 - p_{5}^2 =   p_{2} p_{3}. \\
(\lambda_{\phi} + p_{5}) (\lambda_{\phi} - p_{6}) =  p_{1} p_{3},
\end{align}
\end{subequations}
The last relation is obtained by equating the relation (\ref{invert2}) with 
 (\ref{invert1}). The second is nothing but the first, after the 
$\pi/2$ rotation symmetry $(\vec{b},\vec{m}) \to (\vec{b}^{\perp},\vec{m}^{\perp})$. 
These equation also imply:
 \begin{equation}\label{def-subelastic3BIS}
(\lambda_{\phi} + p_{6}) (\lambda_{\phi} - p_{5}) =  p_{2} p_{4}
\quad ; \quad
(p_{5}-p_{6})^2 = (p_{1}-p_{2}) (p_{3}-p_{4})
\end{equation}

%

It is now simple to check that the ''elastic manifold'' (\ref{manifold}) is preserved by the RG. For the two
first conditions it is straightforward, and for the third one can show and use that:
%
 \begin{equation}
 p_{1}\partial_{l}p_{3} + p_{3}\partial_{l}p_{1} 
 + (\lambda_{\phi}+p_{5}) \partial_{l}p_{6}-(\lambda_{\phi}-p_{6})\partial_{l}p_{5} = 0
 \end{equation}

We can now write the RG equations restricted to this subspace. 
Using the above expressions of the $p_{i}$ in terms of the elastic matrices, we obtain 
the main result of the paper:
\begin{subequations}\label{eq:fullRG-elastic}
\begin{align}
\partial_{l}(c_{11}-c_{66})  = &
-\left( \Gamma_{1} -2 \tilde{\Gamma}_{1} \right)  \frac{2 a_{0}^2}{T}\left(c_{11}-c_{66}\right)^2
\\
& + \left( \Gamma_{2}-2\tilde{\Gamma}_{2} \right)  2 a_{0} |\vec{G}_{1}| \left(c_{11}-c_{66}\right)
  + \left( \Gamma_{3}+2 \tilde{\Gamma}_{3}\right)  \frac12 T |\vec{G}_{1}|^2
\nonumber
\\
\partial_{l}c_{66} = &
-\Gamma_{1} \frac{2 a_{0}^2}{T} c_{66}^2
- \Gamma_{2} 2 a_{0} |\vec{G}_{1}| c_{66} 
 +\Gamma_{3} \frac{|\vec{G}_{1}|^2 T}{2}  
\\
\partial_{l}\gamma = &
-\left( \Gamma_{1} +2 \tilde{\Gamma}_{1} \right)  \frac{2 a_{0}^2}{T}\gamma^2
\nonumber \\
&  + \left( \Gamma_{2}+2\tilde{\Gamma}_{2} \right)  2 a_{0} |\vec{G}_{1}| \gamma
  + \left( \Gamma_{3}-2 \tilde{\Gamma}_{3}\right)  \frac12 T |\vec{G}_{1}|^2 .
\end{align}
\end{subequations}
where the $\Gamma_i$ and $\tilde \Gamma_i$ were defined in (\ref{eq:Gamma}, \ref{eq:TrTM}).
Their explicit calculation and analysis of the resulting equations, in the
specific models, go well beyond this paper. 

\subsection{symmetries}

Let us comment the symmetries of these equations.
They are invariant under the transformation:
\begin{align}
& c'_{11}- c'_{66} = \gamma \\
& c'_{66} = c_{66} \\
& \gamma' = c_{11}-c_{66}. 
\end{align}
which, as noted above, results from invariance under a
$\pi/2$ charge rotation (\ref{sym:rotation}). It means that
if $(c_{11}(l),c_{66}(l),\gamma(l), Y_l[\vec {\bf b}, \vec {\bf m}])$ is
a solution of the RG flow, then $(c'_{11}(l),c'_{66}(l),\gamma'(l), Y_l[\vec {\bf b}, \vec {\bf m}])$
is also a solution. The self-adjoint manifold corresponds to
$K_2=K_4=K_6=0$, i.e. $\gamma=c_{11}-c_{66}$ which corresponds to an
isotropic elastic energy and interaction between charges. It
is a family of conformally invariant VECG. Examples have been studied
in \cite{nelson78} (electric case) and in \cite{boyanovsky91}. 




Similarly, the electromagnetic duality (\ref{sym:duality}) is written as 
\begin{align}
c'_{11}-c'_{66} & = \frac{T^2  |\vec{G}_{1}|^2}{4 a_{0}^2} ~ 1/(c_{11}-c_{66}) \\
c'_{66} & = \frac{T^2  |\vec{G}_{1}|^2}{4 a_{0}^2} ~ 1/c_{66} \\
\gamma' & =  \frac{T^2  |\vec{G}_{1}|^2}{4 a_{0}^2} ~ 1/\gamma . 
\end{align}
It means that
if $(c_{11}(l),c_{66}(l),\gamma(l), Y_l[\vec {\bf b}, \vec {\bf m}])$ is
a solution of the RG flow, then $(c'_{11}(l),c'_{66}(l),\gamma'(l), Y_l[\vec {\bf m}^\perp, \vec {\bf b}^\perp])$
is also a solution. Hence there is a self-dual submanifold invariant by the
flow, defined by:
\begin{align}
c_{11}- c_{66} = c_{66} = \gamma
= T  |\vec{G}_{1}| / (2 a_{0}) \\
Y_l[\vec {\bf b}, \vec {\bf m}] = Y_l[\vec {\bf m}^\perp, \vec {\bf b}^\perp]
\end{align}
In the space of elastic constants this self-dual point forms a ''line'' as
$T$ varies. This manifold is clearly included in the 
''conformal submanifold'' $c_{11}- c_{66} = \gamma$ defined above (it obeys
$K_1=K_3$, $K_2=-K_4=0$, $K_5=K_6=0$ ).

 
\section{Conclusion}

To conclude, we have shown how to derive the RG equations of pinned two dimensional defective solids
from generalized ``elastic'' electromagnetic Coulomb gases with vector charges, defined in 
(\ref{eq:Z-VECG-lattice},\ref{eq:S-VECG-lattice}). These RG equations were obtained 
to lowest order in the charge fugacity, and displayed in full generality in appendix 
 \ref{sec:fullRG}. They involve, in addition to charge fugacities, six elastic 
coefficients (or replica matrices in the disordered case). 
We found that they decouple in two independent sets of simpler equations 
 (\ref{RGeq-p-1}) and (\ref{RGeq-p-2}) which obey two additional symmetry relations.
We found that these general equations exhibit a restriction to only three scaling 
elastic coefficients, corresponding to the initial pinned elastic 
 models, which we showed to be preserved under the RG flow. This
provides our final result : eqs.~(\ref{eq:fullRG-elastic}) which is
still sufficiently general to include all known cases, 
e.g. the scalar electromagnetic Coulomb gas\cite{nienhuis87}, the scalar vector Coulomb gas describing the melting  transition of 2D elastic solids\cite{nelson83b}, together
with various extensions, e.g. the melting transition of pinned 2D solids \cite{carpentier98a}.
Their detailed analysis is the subject of a separate publication.

{\it Acknowledgments :} We thank the Kavli Institute for Theoretical Physics for its hospitality
where the very last stage of this work was completed, and PLD acknowledges support from
ANR grant 05-BLAN-0099-01.

\newpage
\appendix

\section{Two dimensional dislocations}\label{app:dislocations}

 In this appendix, we derive the displacement field corresponding to 
a finite 
density of 2D edge dislocations (and of disclinations) in the presence of a coupling to a substrate. 
We present it here for sake of completeness, and to
clarify the notations used in this paper. 

We  consider a 2D isotropic elastic lattice coupled to a periodic substrate according to 
$H=\frac{1}{2}\int d^{2}\vec{r}~ 2\mu u_{ij}^{2}+\lambda u_{kk}^{2}+\gamma
(\epsilon_{ij}\partial_{i}u_{j})^{2}$. Without dislocations, the phonon 
displacement field ${\bf u}$ is single valued and satisfies
$\epsilon_{ij}\partial _{i} \partial_{j}{\bf u}=0$. Using this property, we can show that 
(apart from boundary terms) 
\[
\int_{\vec{r}} (\epsilon_{ij}\partial_{i}u_{j})^{2} =
2 \int_{\vec{r}} \left(u_{ij}^{2}-u_{kk}^{2} \right)
\]
which implies that the coupling constant to the subtrate $\gamma$ can be
incorporated in new Lam{\'e} coefficients $\tilde{\lambda}=\lambda-2\gamma$ and 
$\tilde{\mu}=\mu+\gamma$ and thus is not a new independant 
elastic constant of the lattice : 
\begin{equation}
\label{app-hamil}
H=\frac{1}{2}\int d^{2}\vec{r}~ 2\mu u_{ij}^{2}+\lambda u_{kk}^{2}+\gamma
\theta^{2} = 
\frac{1}{2}\int d^{2}\vec{r} ~2 (\mu+\gamma) u_{ij}^{2}+ (\lambda-2\gamma)
u_{kk}^{2} 
\end{equation}
 This transformation can also be written as $c_{11}\rightarrow \tilde{c}_{11}=c_{11},
c_{66}\rightarrow \tilde{c}_{66}=c_{66}+\gamma$. As we will see, the appearence of
dislocations breaks this symmetry. 

%

  The local equilibrium condition for the hamiltonian (\ref{app-hamil})
$H=\frac{1}{2} u_{i}*M_{ij}*u_{j}$ reads
\begin{equation}\label{local-equi}
\frac{\partial H}{\partial u_{i}} =0 \Rightarrow  M_{ij} * u_{j} =
 2 \mu~ \partial_{j} u_{ij} +\lambda~ \partial_{i} u_{kk}
+\gamma \epsilon_{ji}\epsilon_{mn}\partial_{j} \partial_{m}u_{n} =0
\end{equation}
Only for non singular fields does the matrix $M_{ij}$ reduce to  :
$M_{ij} ({\bf q})=q^{2}[(2\tilde{\mu}+\tilde{\lambda})P_{ij}^{L}+
\tilde{\mu}P_{ij}^{T}]$ where we have use the modified Lam{\'e}
coefficients introduced above. 

 Since dislocations correspond to topological singularities of the lattice, 
they induce multi-valued displacement fields $u_{i}$. Hence if we want to formulate 
the problem of the determination of their displacement field as a classical elasticity  
problem,  we need to split the displacement field $u_{i}$ into a multi-valued 
part $u^{s}_{i}$ and a smooth component $\tilde{u}_{i}$ :
$u_{i}=u^{s}_{i}+\tilde{u}_{i}$. The field $u^{s}_{i}$ provides the necessary 
multi-valueness : $u^{s}_{i}=b_{i}*\frac{\Phi}{2\pi}$ where
${\bf b} (\vec{r}) =\sum_{\alpha}\delta ({\bf r-r_{\alpha}}) {\bf b}_{\alpha}
$ is the dislocations density. Using $\partial_{i}
\Phi=-\epsilon_{ik}\partial_{k} G$, we find the contribution of $u_{i}^{s}$ to
(\ref{local-equi}) : 
\begin{eqnarray}
&& u_{ij}^{s} = -\frac{1}{4\pi} (b_{j}\epsilon_{ik}+b_{i}\epsilon_{jk})
\partial_{k} G
\label{eq:uijs}\\
&& \Rightarrow  f_{i}^{s}\equiv 
-M_{ij}*u^{s}_{j} =
- 2 \mu~ \partial_{j} u^{s}_{ij} -\lambda~ \partial_{i} u^{s}_{kk}
-\gamma \epsilon_{ji}\epsilon_{mn}\partial_{j} \partial_{m}u^{s}_{n}\\
&&~~~~~~~= 
\frac{1}{2\pi} b_{j} \left( (\tilde{\mu}-2\gamma) \epsilon_{ik}
 \partial_{j} \partial_{k}+ 
(\tilde{\lambda}+2\gamma) \epsilon_{jk} \partial_{i} \partial_{k}\right)G
\end{eqnarray}

 We are now facing a classical elasticity problem consisting of finding the response of an
isotropic 2D lattice under a local force $f_{i}^{s}$ : 
$ M_{ik}*\tilde{u}_{k}=f_{i}^{s}$, which can be inverted in Fourier 
transform as 
\[
\tilde{u}_{i} ({\bf q}) = M_{ij}^{-1} ({\bf q}) f_{j}^{s} ({\bf q})
\]
 with 
 $$M_{ij}^{-1}=\frac{1}{q^{2}} (\frac{1}{2\tilde{\mu}+\tilde{\lambda}}
P_{ij}^{L}+\frac{1}{\tilde{\mu}}P_{ij}^{T})$$ 
and 
$$f_{i}^{s}=-b_{j} ((\tilde{\mu}-2\gamma) \epsilon_{ik}P_{jk}^{L}+
(\tilde{\lambda}+2\gamma) 
\epsilon_{jk} P_{ik}^{L}).$$ 
We end up with a displacement field
$\tilde{u}$ given by 
\begin{equation}
 \tilde{u}_{i} ({\bf q})= -b_{j} \frac{1}{q^{2}}
\left(
\frac{\tilde{\lambda}+2\gamma}{2\tilde{\mu}+\tilde{\lambda}} 
\epsilon_{jk} P_{ik}^{L}
+\frac{\tilde{\mu}-2\gamma}{\tilde{\mu}} \epsilon_{ik}P_{jk}^{L} \right)
\label{eq:utilde}
\end{equation}
 Using the approximate Fourier transform 
\[
\int \frac{d^{2}{\bf q}}{(2\pi)^{2}} e^{i {\bf q}.\vec{r}}
\frac{1}{q^{2}}P_{ij}^{L}=
-\frac{1}{4\pi} \left(\delta_{ij}\ln r
+\frac{r_{i}r_{j}}{r^{2}}-\frac{1}{2}\delta_{ij} +C (\phi)\right) 
\]
 and the relation 
 $\epsilon_{jk} \hat{r}_{i}\hat{r}_{k}= \epsilon_{ik} \hat{r}_{j}\hat{r}_{k}+\epsilon_{ij}$, 
 we obtain the result :
\begin{equation}
\tilde{u}_{i}= \frac{b_{j}}{2\pi} \left( 
\frac{\tilde{\mu}^{2}-\gamma (3\tilde{\mu}+\tilde{\lambda})}
{\tilde{\mu}( 2\tilde{\mu} +\tilde{\lambda})} 
\epsilon_{ij}\ln (r) 
+
\frac{(\tilde{\mu}+\tilde{\lambda})(\tilde{\mu}-\gamma)}
{\tilde{\mu}(2\tilde{\mu} + \tilde{\lambda})} 
\epsilon_{jk}H_{ik} \right). 
\end{equation}
 Thus the total displacement field due to a density of dislocations ${\bf b}$
is $u_{i}=\frac{1}{2\pi} \mathcal{G}_{ij}*b_{j}$ where 
\begin{equation}
\mathcal{G}_{ji} (r)=
\delta_{ij} \Phi (r)+
\frac{\tilde{c}_{66}^2 - \gamma (\tilde{c}_{11}+\tilde{c}_{66})}{\tilde{c}_{11}\tilde{c}_{66}} \epsilon_{ij}\ln (r) +
\frac{(\tilde{c}_{11}-\tilde{c}_{66})(\tilde{c}_{66}-\gamma)}{\tilde{c}_{11}\tilde{c}_{66}} \epsilon_{jk}H_{ik} (r)
\end{equation}
 This expression reduces to the known formula\cite{lubensky96} without any coupling to the substrate $\gamma=0$. 
 We also realize that in the presence of dislocations, this coupling  $\gamma$ can no longer be incorporated into 
 renormalized elastic coupling : its corresponds to a third independent constant. 
 
To obtain the effective interaction between the dislocations, we first express the strain tensor corresponding to a collection 
of dislocations : 
$u_{ij} ({\bf q}) = u_{ij}^{s} ({\bf q}) + \tilde{u}_{ij} ({\bf q})$ where, using (\ref{eq:uijs})
\begin{equation}
u_{ij}^{s} ({\bf q}) = 
i\left( b_{j} \epsilon_{ik}+b_{i}\epsilon_{jk}\right)\frac{q_{k}}{2q^{2}} 
\end{equation}
 and from (\ref{eq:utilde})
\begin{equation}
\tilde{u}_{ij} ({\bf q}) = 
- i \frac{b_{l}}{2 q^2} \left(
\frac{2\tilde{\lambda}+4\gamma}{2\tilde{\mu}+\tilde{\lambda}} 
\epsilon_{lk}q_{i}P_{jk}^{L}+
\frac{\tilde{\mu}-2\gamma}{\tilde{\mu}}
(\epsilon_{jk}q_{i}+\epsilon_{ik}q_{j})P_{lk}^{L} \right). 
\end{equation}
  Now plugging this strain tensor into the elastic energy and using
\begin{align*}
& u_{ij}^{s}({\bf q})u_{ij}^{s}(-{\bf q}) = 
\frac{1}{2 q^2}b_{i}({\bf q})b_{j}({-\bf q}) \left( \delta_{ij} + P_{ij}^{T}\right), 
\\
& \tilde{u}_{ij}({\bf q})\tilde{u}_{ij}(-{\bf q}) = 
\frac{1}{4 q^2} b_{i}({\bf q})b_{j}({-\bf q}) \left[ 
\left( \frac{2\tilde{\lambda}+4\gamma}{2\tilde{\mu}+\tilde{\lambda}}  \right)^2 P_{ij}^{T}
+2 \left(\frac{\tilde{\mu}-2\gamma}{\tilde{\mu}} \right)^2 P_{ij}^{L}
\right]
\\
& \tilde{u}_{ij}({\bf q})u^{s}_{ij}(-{\bf q}) = 
-\frac{1}{2 q^2} \left(\frac{\tilde{\mu}-2\gamma}{\tilde{\mu}} \right) 
b_{i}({\bf q})b_{j}({-\bf q}) P_{ij}^{L}
\\
& u_{kk}({\bf q})u_{kk}(-{\bf q}) = 
\frac{1}{4q^2} 
\left(2-\frac{2\tilde{\lambda}+4\gamma}{2\tilde{\mu}+\tilde{\lambda}} \right)^2 
b_{i}({\bf q})b_{j}({-\bf q}) P_{ij}^{T}
\end{align*}  
  we obtain the desired result :  
\begin{align}\label{H-disloc}
H_{b/b} &= 
\int_{q}
\frac{1}{2q^2} b_{i}({\bf q})b_{j}({-\bf q}) 
\left[
\frac{4 \gamma^2}{\tilde{c}_{66}} P_{ij}^{L}
+
\frac{4 (\tilde{\mu}(\tilde{\mu}+\tilde{\lambda}) + \gamma^2)}{2\tilde{\mu} +\tilde{\lambda}} 
 P_{ij}^{T} 
 \right]
 \\
 &=\int_{q}
\frac{1}{2q^2} b_{i}({\bf q})b_{j}({-\bf q}) 
 \left[
\frac{4 \gamma^2}{\tilde{\mu}} P_{ij}^{L}
+
\frac{4c_{66} (c_{11}-c_{66})+4\gamma^2}{c_{11}}
 P_{ij}^{T} 
 \right]
\end{align}
Note that another method to obtain this interaction,  incorporating in particular the contribution of disclinations,
 is to use the so called Airy functions\cite{lubensky96}. However it does not
provide the displacement field, necessary in the present case.
\section{Renormalization Group Equations for the Full Model}
\label{sec:fullRG}

In this appendix, we present the RG equations for the full VECG. In these expression, the couplings $K_{i}$, $ \Gamma_i$ and $\tilde \Gamma_i$ 
 are commuting replica matrices.  

\begin{subequations}  \label{eq:RGfull}
\begin{align}
\partial_{l} K_{1}^{ab} &= 
-2\pi \Gamma_{1}             \left(  K_{1}^2 + K_{2}^2 \right)  
+4\pi \tilde{\Gamma}_{1}   K_{1}K_{2}
\nonumber
\\ &
+4\pi \Gamma_{2}           \left( K_{1}K_{5}  -K_{2}K_{6} \right)
-4\pi  \tilde{\Gamma}_{2} \left( K_{1}K_{6} - K_{2}K_{5} - K_{2}   \frac{\lambda_{\phi}}{(2\pi)} \right) 
\nonumber
\\ &
+2\pi \Gamma_{3}             \left(   K_{5}^2 +K_{6}^2  +  \left( \frac{\lambda_{\phi}}{(2\pi)}\right)^2 \right)
-4\pi \tilde{\Gamma}_{3}   \left(     K_{6}  \frac{\lambda_{\phi}}{(2\pi)} 
                                         K_{5}K_{6}\right)
\end{align}
\vspace{-1cm}
\begin{align}
\partial_{l} K_{2}^{ab} &= 
-4\pi \Gamma_{1}           K_{1}K_{2} 
+2\pi \tilde{\Gamma}_{1} \left( K_{1}^2-K_{2}^2\right) 
\nonumber
\\ &
+4\pi  \Gamma_{2} 		\left( K_{2}K_{5}-K_{1}K_{6}\right)
+4\pi \tilde{\Gamma}_{2} \left( K_{1}K_{5} -K_{2}K_{6} +K_{1}\frac{\lambda_{\phi}}{(2\pi)} \right) 
\nonumber
\\ &
-4\pi \Gamma_{3} 		K_{5}K_{6} 
+2\pi \tilde{\Gamma}_{3} \left( K_{5}^2 + K_{6}^2 + 2K_{5} \frac{\lambda_{\phi}}{(2\pi)} 
						+ \left(\frac{\lambda_{\phi}}{(2\pi)}\right)^2\right)
\end{align}
\vspace{-1cm}
\begin{align}
\partial_{l} K_{3}^{ab} &= 
+2\pi \Gamma_{1}           \left(K_{5}^2+K_{6}^2+\left(\frac{\lambda_{\phi}}{(2\pi)}\right)^2 \right)
+4\pi \tilde{\Gamma}_{1} \left( K_{5}K_{6}) 
					- K_{6} \frac{\lambda_{\phi}}{(2\pi)} \right) 
\nonumber
\\ &
+4\pi  \Gamma_{2} 		\left( K_{3}K_{5} +K_{4}K_{6} \right)
-4\pi \tilde{\Gamma}_{2} \left( K_{4}K_{5}+K_{3}K_{6}
					-K_{4}\frac{\lambda_{\phi}}{(2\pi)}\right) 
\nonumber
\\ &
-2\pi \Gamma_{3} 		\left( K_{3}^2+K_{4}^2\right)
+4\pi \tilde{\Gamma}_{3} K_{3}K_{4} 
\end{align}
\vspace{-1cm}
\begin{align}
\partial_{l} K_{4}^{ab} &= 
+4\pi \Gamma_{1}           K_{5}K_{6} 
+2\pi \tilde{\Gamma}_{1} \left( \left(\frac{\lambda_{\phi}}{(2\pi)}\right)^2 
					-2K_{5}\frac{\lambda_{\phi}}{(2\pi)}
					+K_{5}^2+ K_{6}^2  \right) 
\nonumber
\\ &
+4\pi  \Gamma_{2} 		\left( K_{4}K_{5}+ K_{3}K_{6}\right)
-4\pi \tilde{\Gamma}_{2} \left( K_{3}K_{5} -  K_{3} \frac{\lambda_{\phi}}{(2\pi)}
					+ K_{4}K_{6}\right) 
\nonumber
\\ &
-4\pi \Gamma_{3} 		K_{3}K_{4}
+2\pi \tilde{\Gamma}_{3} \left( K_{3}^2-K_{4}^2 \right)
\end{align}
\begin{align}
\partial_{l} K_{5}^{ab} &= 
2\pi \Gamma_{1}           \left(K_{2}K_{6}-K_{1}K_{5} \right)
+2\pi \tilde{\Gamma}_{1} \left( - \frac{\lambda_{\phi}}{(2\pi)} K_{2} - K_{1}K_{6}+K_{2}K_{5} \right) 
\nonumber
\\ &
+2\pi  \Gamma_{2} 		\left(- \left(\frac{\lambda_{\phi}}{(2\pi)} \right)^2 +K_{5}^2+K_{6}^2
					- K_{1}K_{3}+K_{2}K_{4} \right)
\nonumber
\\ &
-2\pi \tilde{\Gamma}_{2} \left( K_{2}K_{3}-K_{1}K_{4}+2K_{5}K_{6}\right) 
\nonumber
\\ &
-2\pi \Gamma_{3} 		\left(K_{3}K_{5}+K_{4}K_{6} \right)
+2\pi \tilde{\Gamma}_{3} \left( K_{4}\frac{\lambda_{\phi}}{(2\pi)} +K_{3}K_{6}+K_{4}K_{5} \right)
\end{align}
\vspace{-1cm}
\begin{align}
\partial_{l} K_{6}^{ab} &= 
2\pi \Gamma_{1}           \left(K_{2}K_{5}-K_{1}K_{6} \right)
+2\pi \tilde{\Gamma}_{1} \left(K_{1}\frac{\lambda_{\phi}}{(2\pi)} -K_{1}K_{5}+K_{2}K_{6} \right)
 \nonumber
\\ &
+2\pi \Gamma_{2} 		\left( K_{2}K_{3}-K_{1}K_{4}+2K_{5}K_{6}\right)
\nonumber
\\ &
-2\pi \tilde{\Gamma}_{2} \left( -\left( \frac{\lambda_{\phi}}{(2\pi)}\right)^2 + K_{5}^2  +K_{6}^2 
					-K_{1}K_{3}+K_{2}K_{4} \right)
\nonumber
\\ &
-2\pi \Gamma_{3} 		\left(K_{4}K_{5}+K_{3}K_{6} \right)
+2\pi \tilde{\Gamma}_{3} \left( K_{3}\frac{\lambda_{\phi}}{(2\pi)}+K_{3}K_{5}+K_{4}K_{6}
					\right)
\end{align}
\end{subequations}


%
%

\end{document}